\documentclass[iop]{emulateapj}

\usepackage{natbib}
\newcommand{\Nsysdisc}{4}

\newcommand{\Kepler}{{\sl Kepler}\ }

\newcommand{\Kepa}{Kepler-29}  % koi738
\newcommand{\Kepb}{Kepler-30}  % koi806
\newcommand{\Kepc}{Kepler-31}  % koi935
\newcommand{\Kepd}{Kepler-32}  % koi952
\newcommand{\be}{\begin{equation}}
\newcommand{\ee}{\end{equation}}
\newcommand{\bea}{\begin{eqnarray}}
\newcommand{\eea}{\end{eqnarray}}
\shorttitle{Confirmation of \Nsysdisc~Multiple Planet Systems}
\shortauthors{Fabrycky et al.}

\begin{document}
\title{
Transit Timing Observations from \Kepler:  
IV. Confirmation of \Nsysdisc~Multiple Planet Systems by Simple Physical Models
}
\author{
Daniel C. Fabrycky\altaffilmark{1,2},  % wrote it, did dynamical calculations
%
% Top-flight contributors:
Eric B. Ford\altaffilmark{3}, % drove general effort, advocated for FOP observations, comparison to other method.  Ford contributed to the transit time variations analysis and interpretation.
Jason H. Steffen\altaffilmark{4},   % comparison to other method
% Major from TTV Group
Jason F. Rowe\altaffilmark{5,6},    % KOIs, bulk TTs
Joshua A. Carter\altaffilmark{7,2},   % Special TTs: did the times for 806b.
Althea V. Moorhead\altaffilmark{3}, % assembling FOP info and lightcurve analysis
% Other
Natalie M. Batalha\altaffilmark{8}, % selecting and vetting candidates
William J. Borucki\altaffilmark{6}, % develop the Mission that obtained the data 
Steve Bryson\altaffilmark{6},  % SOC Centroids
Lars A. Buchhave\altaffilmark{18,19},  % CfA Spectra group
Jessie L. Christiansen\altaffilmark{5,6},  % SOC support (through Jenkins)
David R. Ciardi\altaffilmark{16},
William D. Cochran\altaffilmark{17}, % McDonald.    obtained and processed the McDonald 2.7m recon spectrum of KOI-952, and processed the McDonald 2.7m recon spectrum of KOI-806
Michael Endl\altaffilmark{17},  % McDonald.  I observed the recon spectrum for K00806 in July 2010 with the 2.7 m telescope at McDonald Observatory.
Michael N. Fanelli\altaffilmark{12},% SOC support (through Jenkins)
Debra Fischer\altaffilmark{13},  % SME analysis of 806's spectrum.
Francois Fressin\altaffilmark{7}, % Blender
John Geary\altaffilmark{7}, % development of Kepler photometer electronics, Keplercam for KIC and followup spectral typing, echelle spectrograph at SAO for follow-up.
Michael R. Haas\altaffilmark{6}, % Science Office.
Jennifer R. Hall\altaffilmark{9}, % SOC support (through Jenkins)
Matthew J. Holman\altaffilmark{7},  % general consul
Jon M. Jenkins\altaffilmark{5,6},  % SOC TT,  the set of light curves containing the planetary signatures.
David G. Koch\altaffilmark{6}, % everything.  I contributed to the concept, design, development and testing of the Kepler mission
David W. Latham\altaffilmark{7}, % CfA Spectra group, TTV telecons
Jie Li\altaffilmark{5,6},  % SOC support (through Jenkins)
Jack J. Lissauer\altaffilmark{6},   % general consul.  My contribution included work on determining criteria for considering probabilistic evidence for planethood as confirmation.
Philip Lucas\altaffilmark{10}, % UKIRT Images for FP rejection.   My contribution was the UKIRT J band images.
Geoffrey W. Marcy\altaffilmark{11}, % Keck/Lick
Tsevi Mazeh\altaffilmark{20}, % Performed autocorrelation analysis for starspot periods.
Sean McCauliff\altaffilmark{6}, % SOC
Samuel Quinn\altaffilmark{7}, % performed the initial analysis of the reconnaissance spectroscopy.
Darin Ragozzine\altaffilmark{7},    % general consul.  Providing consultation on the methods useful for confirming these multi-transiting systems.
Dimitar Sasselov\altaffilmark{7}, % I took part in the discussions about the method and the implications of the results.
Avi Shporer\altaffilmark{14,15} % FTN; I collected imaging for 3 of the 4 targets presented in the paper: KOI-738, KOI-935 and KOI-952, all in the SDSS-r. I examined all images for non-KIC targets near the target, as such objects are not accounted for by the contamination factor.
}
\altaffiltext{1}{Department of Astronomy and Astrophysics, University of California, Santa Cruz, Santa Cruz, CA 95064, USA}
\altaffiltext{2}{Hubble Fellow}
\altaffiltext{3}{Astronomy Department, University of Florida, 211 Bryant Space Sciences Center, Gainesville, FL 32111, USA}
\altaffiltext{4}{Fermilab Center for Particle Astrophysics, P.O. Box 500, MS 127, Batavia, IL 60510, USA}
\altaffiltext{5}{SETI Institute, Mountain View, CA, 94043, USA}
\altaffiltext{6}{NASA Ames Research Center, Moffett Field, CA, 94035, USA}
\altaffiltext{7}{Harvard-Smithsonian Center for Astrophysics, 60 Garden Street, Cambridge, MA 02138, USA}
\altaffiltext{8}{San Jose State University, San Jose, CA 95192, USA}
\altaffiltext{9}{Orbital Sciences Corporation/NASA Ames Research Center, Moffett Field, CA 94035, USA}
\altaffiltext{10}{Centre for Astrophysics Research, University of Hertfordshire, College Lane, Hatfield, AL10 9AB, England}
\altaffiltext{11}{University of California, Berkeley, Berkeley, CA 94720}
\altaffiltext{12}{Bay Area Environmental Research Institute/NASA Ames Research Center, Moffett Field, CA 94035, USA}
\altaffiltext{13}{Astronomy Department, Yale University, New Haven, CT, USA ; Department of Physics and Astronomy, San Francisco State University, San Francisco, CA, USA}
\altaffiltext{14}{Las Cumbres Observatory Global Telescope Network, 6740 Cortona Drive, Suite 102, Santa Barbara, CA 93117, USA}
\altaffiltext{15}{Department of Physics, Broida Hall, University of California, Santa Barbara, CA 93106, USA}
\altaffiltext{16}{NASA Exoplanet Science Institute/Caltech, Pasadena, CA 91126, USA}
\altaffiltext{17}{McDonald Observatory, The University of Texas, Austin TX 78730, USA}
\altaffiltext{18}{Niels Bohr Institute, University of Copenhagen, DK-2100 Copenhagen, Denmark}
\altaffiltext{19}{Centre for Star and Planet Formation, Natural History Museum of Denmark, University of Copenhagen, DK-1350 Copenhagen, Denmark }
\altaffiltext{20}{School of Physics and Astronomy,Raymond and Beverly Sackler Faculty of Exact Sciences, Tel Aviv University, Tel Aviv 69978, Israel}
\email{daniel.fabrycky@gmail.com}

\begin{abstract}
Eighty planetary systems of two or more planets are known to orbit stars other than the Sun.  For most, the data can be sufficiently explained by non-interacting Keplerian orbits, so the dynamical interactions of these systems have not been observed.  Here we present \Nsysdisc~sets of lightcurves from the \Kepler spacecraft, which each show multiple planets transiting the same star.  Departure of the timing of these transits from strict periodicity indicates the planets are perturbing each other: the observed timing variations match the forcing frequency of the other planet.  This confirms that these objects are in the same system.  Next we limit their masses to the planetary regime by requiring the system remain stable for astronomical timescales.  Finally, we report dynamical fits to the transit times, yielding possible values for the planets' masses and eccentricities.  As the timespan of timing data increases, dynamical fits may allow detailed constraints on the systems' architectures, even in cases for which high-precision Doppler follow-up is impractical. 
\end{abstract}

\keywords{planetary systems; stars: individual (KID 10358759 / KOI-738 /  \Kepa;  KID 3832474 / KOI-806 / \Kepb; KID 9347899 / KOI-935 /  \Kepc, KID 9787239 / KOI-952 /  \Kepd); planets and satellites: detection, dynamical evolution and stability; methods: statistical}

\section{Introduction} 
\label{secIntro}

%\subsection{Kepler}
%
% Kepler mission
%
So far, 170 systems with more than one transiting planet candidate have been discovered by \Kepler \citep{2011Borucki}, enabling simple dynamical models for an astounding variety of planetary systems \citep{2011Lissauerb}.  In particular, two very important dynamical quantities, the planetary period and phase, are measured to high precision by {\sl Kepler}.  If we further assume, as is true for the Solar System, that the planets orbit in the same direction, in nearly the same plane, and on nearly circular orbits, then we have a fiducial model that specifies the positions of the planets as a function of time.  However, planets do not follow independent Keplerian orbits; instead planets interact with each other gravitationally.  These small perturbations to the orbits affect the transit times, which can be calculated via numerical simulations.  If the calculated transit timing variations match essential aspects of the observed data, then we obtain independent confirmation that two objects are orbiting the same star. 

In this paper, we develop simple physical models and compare them to \Kepler data to confirm \Nsysdisc~planetary systems, \Kepa-32.  Instead of exhaustively modeling the systems, as has been our practice up until now \citep{2010Holman, 2011Lissauera, 2011Cochran}, we content ourselves with providing upper limits to the planetary masses, based on the principle that the system is extremely unlikely to be dynamically unstable on timescales much less than the age of the star.  Thus, we infer planetary masses for 9 objects in these 4 systems, and deem them planets on this basis.  This paper is being published contemporaneously with two companion papers, which explore anticorrelated transit timing variations for the confirmation of pairs of planets, by \cite{2012Ford} (hereafter Paper II) and \cite{2012Steffen} (hereafter Paper III), which together with this paper form a catalog of $10$ confirmed systems of multiple-transiting planets.  

The organization of this paper is as follows.  Section~\ref{sec_method} introduces the target stars, describes the transit timing data, and discusses how we assess whether the transit timing signal varies on the theoretically-expected timescale.   Section~\ref{sec_results} applies that methodology to confirm \Nsysdisc~planetary systems, and section~\ref{stellar_props} discusses the identification of their host stars.  Section~\ref{sec:Masses} is devoted to constraining the masses of the planets, primarily by dynamical stability, but also via modeling the transit timing signal itself.  Section~\ref{sec:discuss} closes with a discussion of these results and a perspective about the future of transit-timing confirmation of multiply-transiting systems with {\sl Kepler}.

\section{Methods}
\label{sec_method}

\subsection{Determination of Stellar Parameters}

Identifying properties of the host stars studied here are in Table~\ref{tabStars}, e.g., their names in various catalogs.

For stars \Kepa, \Kepb, and \Kepd, we obtained spectra from several different telescopes as part of the \Kepler team's regular ground-based follow-up program.   Two independent analyses were performed, a standard analysis using `Spectroscopy Made Easy' \citep{1996Valenti,2005Valenti}, as well as a newly-formulated analysis called SPC (Buchhave et al. in prep.), to obtain the parameters $T_{\rm eff}$, $\log g$, [Fe/H], and when possible $v \sin i$.  For \Kepc\ we did not obtain a spectrum, but instead adopted these values from the \Kepler input catalog (KIC; \citealt{2011Brown}) -- i.e., based on color photometry -- with generous error bars.  From these values we performed a Bayesian analysis using Yonsei-Yale stellar isochrones \citep{2001Yi} to extract best-fit and $68\%$ confidence regions on $M_\star$ and $R_\star$ (e.g. \citealt{2004Pont,2007Takeda}).  All these values are reported in Table~\ref{tabStarsProps}.  

\subsection{Measurement of Transit Times and Parameters}

The data we use for this study are long-cadence lightcurves from quarters 1 through 6 (and additionally for \Kepc\ only, quarters 7 and 8), available at the Multimission Archive at STScI (MAST\footnote{http://archive.stsci.edu/kepler/}).  

For most of the sample, we used timing results from the general-purpose routines of J. Rowe, previously described in \cite{2011Ford}~[Paper I].  For \Kepb b, extremely large variation in the transit times caused the general-purpose algorithm to fail (a linear model for the times of transit incorrectly predicts some observed times by over half a day, see section~\ref{sec:806}).  Thus, around each transit of this planet, we fit model transit curves \citep{2002MA} and minimized the sum-of-squared residuals, the standard $\chi^2$ function.  We sampled $\chi^2$ over a grid of timing offsets, and fit a parabola to the region within $\Delta \chi^2\leq7$ of the minimum value.  That parabola thus has a width which is more stable to noise properties than the local curvature is, and we adopt its minimum as the transit time and its width (where $\Delta \chi^2=1$) as the error bar.

The transit times for all the planets analyzed herein are given in Table~\ref{tabTTs}.

We also fit transit curves for all our candidates to determine best-fit parameters and errors.  The first step was to detrend the lightcurve; we masked out each transit and fit a polynomial over timescales of $1000$ minutes and divided the whole lightcurve by it, to obtain a detrended, normalized lightcurve $f$.  Second, we defined a contamination-corrected lightcurve via $f_{\rm corr} = (f-c)/(1-c)$, where $c$ is the fractional amount of light leaking into the aperture from known nearby stars, the ``contamination'' reported on MAST for each target for each quarter.  Third, we assembled a phased lightcurve by subtracting the previously-measured transit time from each transit\footnote{For the very small candidates KOI-935.04 and KOI-952.05, individual transit times were not reliably measured (hence they are not included in Table~\ref{tabTTs}), so we used a linear ephemeris for fitting their transit parameters, and held their impact parameters $b$ fixed at 0 and their limb-darkening coefficients $u$ fixed at $0.5$.}.  We used only non-overlapping transits, for which another planet's transit center was not within 5 hours\footnote{We also discarded \Kepc c's transits near $t-245000 = 16.8$ and $400.5$, as these had corrupted data, apparently due to detrending problems near gaps.}.  To construct a flux value of the long cadence (29.4 min), we took 20 samples evenly spaced over that timespan (using {\sc occultsmall}; \citealt{2002MA}) and averaged them.  Finally, we fit the following four parameters: radius ratio $R_p/R_\star$,  duration $T_{\rm dur}$, scaled impact parameter $b$, and linear limb-darkening coefficient $u$.  The duration is defined as the time between mid-ingress and mid-egress, when the center of the planet passes over the limb of the star.  We report these parameters in Table~\ref{tabPlanets}.  

\begin{figure}
\plotone{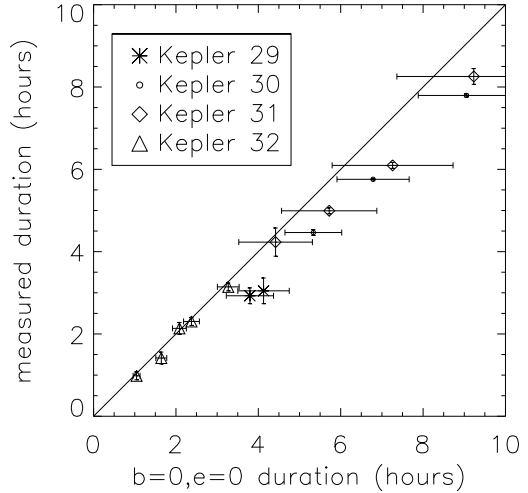}
\caption{ Computed and measured durations of the planet candidates in our sample.  The line shows 1:1.  The data come close to it, providing a useful check on our interpretation of planetary systems orbiting the targets stars.  That the data fall slightly below the line could mean either the stellar parameters or the assumed edge-on and circular orbits are not perfectly correct.  \vspace{0.4 in} }
\label{durations}
\end{figure}

As a check on these stellar and planetary parameters, and as a first indication that all the planet candidates in each system are transiting the target stars, we computed for each planet candidate the transit duration that would be obtained by an orbit that is edge-on ($b=0$) and circular ($e=0$).  These values are plotted against the measured values in Figure~\ref{durations}.  We see only a little evidence that these computed values are slightly larger than the measured ones, an indication that either (i) the computed stellar radii are too large or stellar masses are too small, or (ii) the planets' orbits are systematically seen near pericenter of moderately eccentric orbits, or --- most probably --- (iii) the planets' orbits are not quite edge-on ($b \neq 0$). 

\subsection{Dynamical Expectations and their Observational Confirmation}

We begin N-body integrations near the center of the dataset ( t [BJD] = 2455220 ), using coplanar, circular orbits that match the observed periods and transit epochs, and using nominal masses $M_p = M_\oplus (R_p/R_\oplus)^{2.06}$.  This set of physical models was introduced by \cite{2011Lissauerb}, and it helps us diagnose the most prominent physical interactions.  For systems of more than two planets, we break the problem into planet pairs, in an effort to probe the individual interactions.   First, we calculate transit times as described by \cite{2010Fabrycky} from the beginning of the dataset until 10 years later.  Next, the deviations of those simulated transit times from a linear ephemeris calculated on those times (i.e., the Simulated minus Calculated diagram, $S-C$) is Fourier analyzed to find the dominant frequency \citep{1982S,2009ZK}, which are given in Table~\ref{tabPlanetPairs}. 

We test for variations at this frequency in the observed timing (i.e., the Observed minus Calculated diagram, $O-C$).   The frequency is already determined theoretically (above), but we allow amplitude and phase of the sinusoid and a constant offset to vary freely to fit the data.  This approach contrasts to a periodogram search, which tests many frequencies.  By fixing this important parameter rather than scanning over it, our method is more sensitive than a periodogram search at finding variations at the predicted frequency.  The reason we treat the phase as a free parameter is that it is expected to have a complex dependence on the eccentricities of both the perturbing planet and the perturbed planet.  That is, we cannot rely on the values from a model which assumes circular orbits.  The same is true for amplitude, which further depends on the masses of the planets, which are not known at this initial stage of modeling.  In contrast, the dominant frequency from those models should be present in the data, although perhaps no longer dominant, even if both planets are eccentric or of a mass different than assumed.  This test for a TTV periodicity at the predicted frequency may not yield a significant detection if (a) eccentricity significantly changes the character of the variations, or (b) the transit variations are dominated by a non-transiting planet.  In these cases, large variations may be present, but have little power at the predicted frequency, so more data-driven methods (papers II and III) are more useful.   However, we note that for the systems presented in those papers, the dominant timescale of variation has a simple physical interpretation and is captured by the nominal models described in this paper. 

The amplitude of the best-fitting sinusoid at the theorized frequency is recorded, and it is compared to the same calculation for $10^5$ realizations of the observational $O-C$ data with scrambled entries.  These scrambled datasets should not contain the signal, but they will contain the same distribution of noise (which is possibly non-Gaussian, with outliers).   Our criterion for confirmation is that fewer than $100$ of these realizations yield larger amplitudes than the real data, i.e., a False Alarm Probability (FAP) of $<10^{-3}$.  Note that our method does not require that \emph{both} planets show a TTV signal with such a low FAP.  Even if the transits of one planet are at the limit of detectability, as long as we can establish its orbital period and phase, we have what we need to estimate its dynamical effect on the other planet, whose TTVs must behave as expected.  The FAP statistics are reported in Table~\ref{tabPlanetPairs} for the systems confirmed principally via this method ( i.e., those of Tables~\ref{tabStars}-\ref{tabPlanets}).  Moreover, we present small FAPs reconfirming Kepler-9bc \citep{2010Holman}, Kepler-18cd \citep{2011Cochran}, and nearly all of the systems presented in papers II and III, as these systems show anticorrelated TTVs at the timescale predicted by the nominal models.

%\epsscale{0.7}

\section{Results}

\subsection{Confirmation of \Kepler Planets}
\label{sec_results}
%\subsection{Systems with Significantly Anti-Correlated TTVs}
\label{secSysConfirm}

Here we describe the \Nsysdisc~planetary systems confirmed principally by this method.   Where interesting comparisons can be made to other methods (paper II and paper III), we do so.  Table~\ref{tabStars} gives the catalog properties and Table~\ref{tabStarsProps} gives the physical properties of the host stars detailed in this paper. 

\subsubsection{ KOI-738 = \Kepa}

\begin{figure*}
\plotone{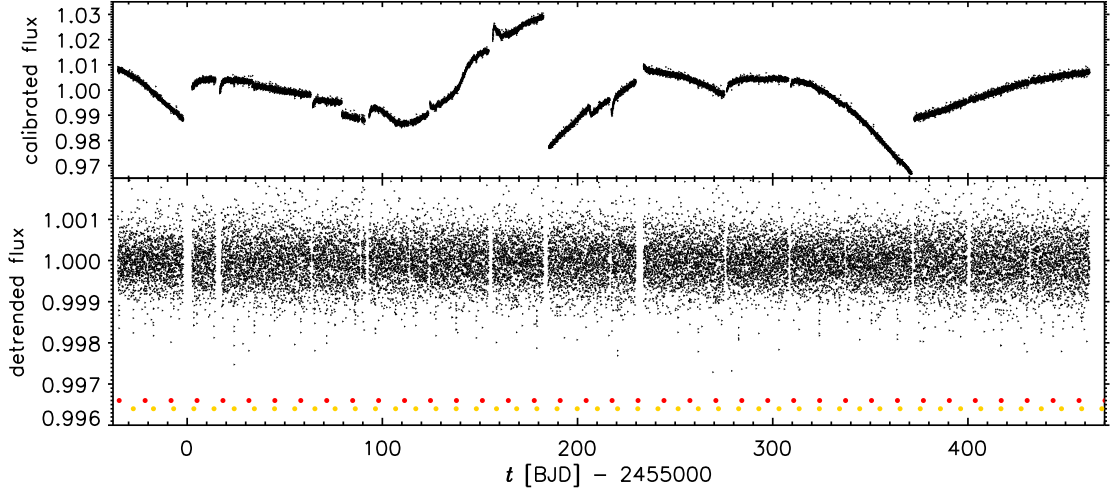}
\caption{\Kepa\ lightcurves.  \emph{Upper panel:} the quarter-normalized, calibrated \Kepler photometry (PA); \emph{lower panel:}  the detrended, normalized flux.  The transit times of each planet are indicated by dots at the bottom of each panel. \vspace{0.4 in} }
\label{738lc}
\end{figure*}

\begin{figure}
\plotone{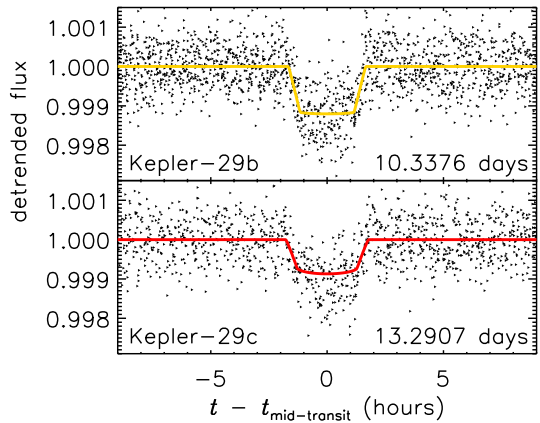}
\caption{\Kepa\ phase curves based on detrended, normalized flux.  Each transit is shifted to its measured midtime, and transits with midtimes within 5 hours of a different planet's midtime are excluded from both this plot and from the model fits.  Overplotted are transit models, box-car smoothed to the 29.4~minute cadence.  The colors correspond to the dots of figure~\ref{738lc}. \vspace{0.4 in}}
\label{738phasecurve}
\end{figure}

The lightcurve of \Kepa\ is plotted in Figure~\ref{738lc}, and the phase curves of the two Neptune-size planets is plotted in Figure~\ref{738phasecurve}.  Their periods of $10.3376(2)$~days and $13.2907(4)$~days form a ratio $1.28566 \pm 0.00005$, which lies right at a 9:7 ($=1.28571$) orbital resonance.  Being fractionally closer than $8\times10^{-5}$ of a second-order resonance in this region would happen only 0.5\% of the time by chance, and similar for other resonant orders.  Being so close to a low-order resonance --- consistent with dynamical libration --- suggests dynamical interaction helped to establish their orbital architecture.  Thus, from the period ratio alone, we have an indication that these objects are in the same system.  

\begin{figure*}
\plotone{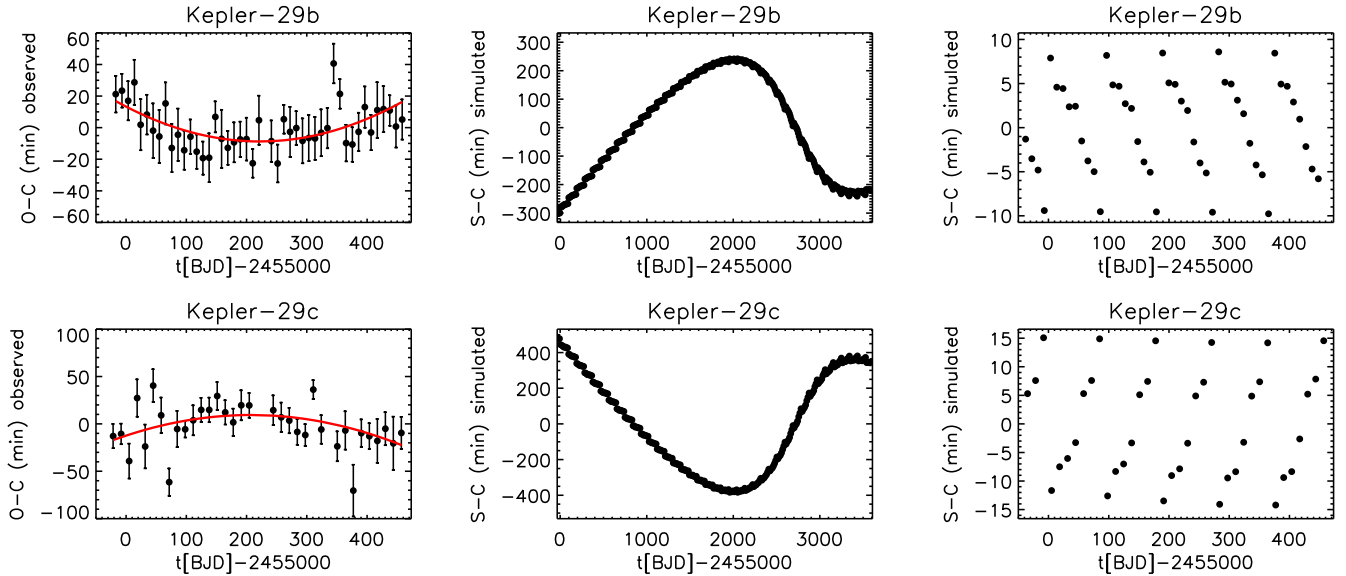}
\caption{ Timing diagrams of \Kepa.  In the left hand panels, the data are plotted.  In red is the best-fitting sinusoid of the theoretically-determined dominant perturbation frequency.  From the amplitude of this sinusoid we determine the significance of the TTV detection.  In the middle panels a nominal model, in which the planets are presumed to start out on circular, coplanar orbits, shown for the decade-span on which we determine the dominant perturbation frequency.  Although this is a poor fit to the data, it illustrates that a very long-timescale and large-amplitude variation is expected due to the planets' orbits being right on top of a resonance with each other.  Also, a very much smaller variation with a $93$~day period is expected.  In the right-hand panels, we have zoomed in on the first part of that signal; these variations are marginally detected in the data. \vspace{0.4 in} }
\label{ot738}
\end{figure*}

Furthermore, since the ratio is not significantly offset from the resonance, the nominal integrations result in an extremely long-timescale variation in $O-C$ (fig.~[\ref{ot738}]), even slightly longer than the sample 10-year integration used to find it.  Therefore, on the timescale of our data, the predicted sinusoid would look only like a quadratic in $O-C$, as is observed (figure~\ref{ot738}, left-hand side).  The data for the shorter-period planet (\Kepa b) indicate this quite robustly ($FAP\simeq 0.02\%$) and hint at oppositely-directed curvature for the longer-period planet (\Kepa c has $FAP \simeq 1.5\%$).  On the basis of the former we confirm the system really is composed of two planets. 

\begin{figure}
\plotone{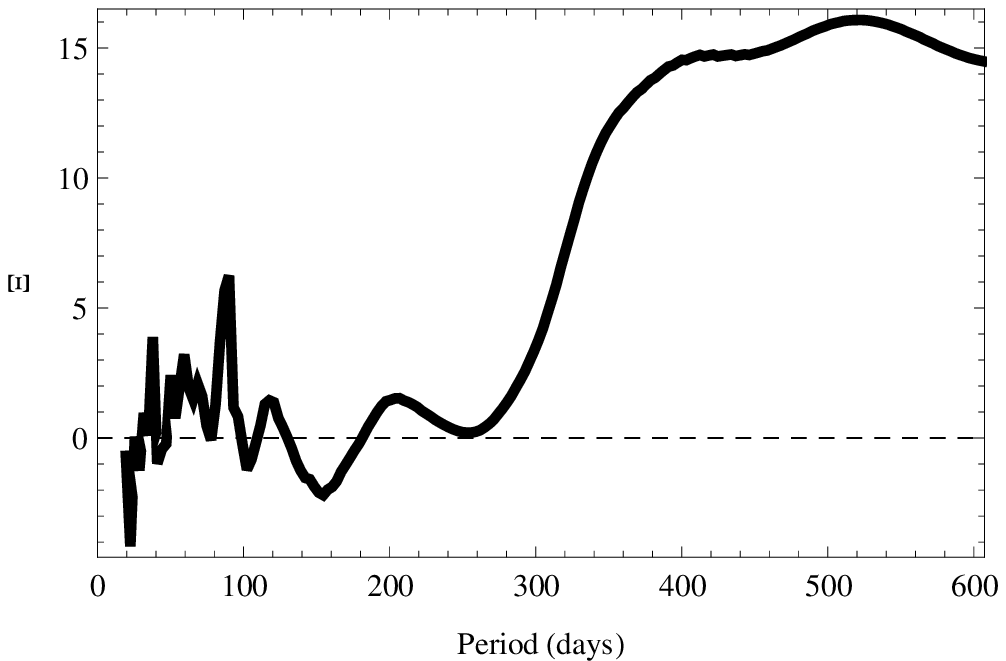}
\caption{ Cross-correlation statistic (paper III) as a function of TTV period for \Kepa\ b/c.  The dominant variation is of long period (as in Figure~\ref{ot738}), and a secondary and only marginally significant peak occurs at $\sim90$days. \vspace{0.4 in} }
\label{steffen738}
\end{figure}

However, a parabolic signature is not specific enough to uniquely specify the other planet as the perturber, so we carried the investigation one more step.  The nominal prediction for year-timescale $O-C$ (figure~\ref{ot738}, right-hand side) is not dominated by a parabola, but is instead a ``chopping'' signal on the frequency at which the planets come back to the same configuration, 
\begin{equation}
P_{ttv} = 9 \times 10.3376 d = 7 \times 13.2907 d = 93 d. \label{eq:ttv738}
\end{equation}  
Thus, we search for that timescale.  After subtracting the curvature models shown in figure~\ref{ot738} (left-hand side), we computed the amplitude of the best-fitting sinusoid at the period of equation~\ref{eq:ttv738}.  For \Kepa b and \Kepa c, the amplitude was 5 and 12 minutes, respectively, quite similar to that predicted by figure~\ref{ot738}.  Then we scrambled the data to find a FAP: it was $0.6\%$ and $1.1\%$ for the two planets, respectively.   To emphasize this point, we plot in Figure~\ref{steffen738} the cross-correlation statistic defined in Paper III: a spike is seen at $\sim90$~days, at positive values of $\Xi$ (i.e., anticorrelation at the predicted timescale).  Therefore, the signal at the expected chopping timescale is a third, albeit modestly significant, indication that these two objects are really planets in the same system.

\epsscale{1.0}

%\epsscale{0.8}

\epsscale{1.2}

\epsscale{1.1}

%\clearpage 

\subsubsection{ KOI-806 = \Kepb} \label{sec:806}

\begin{figure*}
\plotone{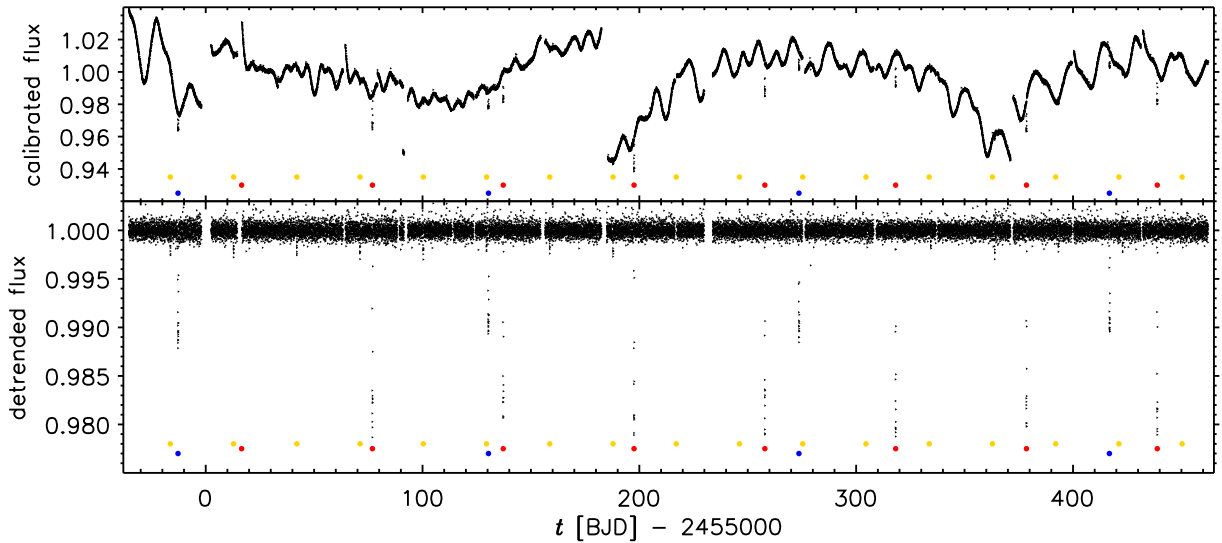}
\caption{\Kepb\ / KOI-806 lightcurves.   \emph{Upper panel:} quarter-normalized calibrated lightcurves; the stellar rotation period of $15.25 \pm 0.25$~days is manifested as out-of-transit variations caused by spots.   \emph{Bottom panel:} Detrended, normalized flux.  The transit times of each planet are indicated by dots at the bottom of each panel. \vspace{0.4 in}  }
\label{806lc}
\end{figure*}

\begin{figure}
\plotone{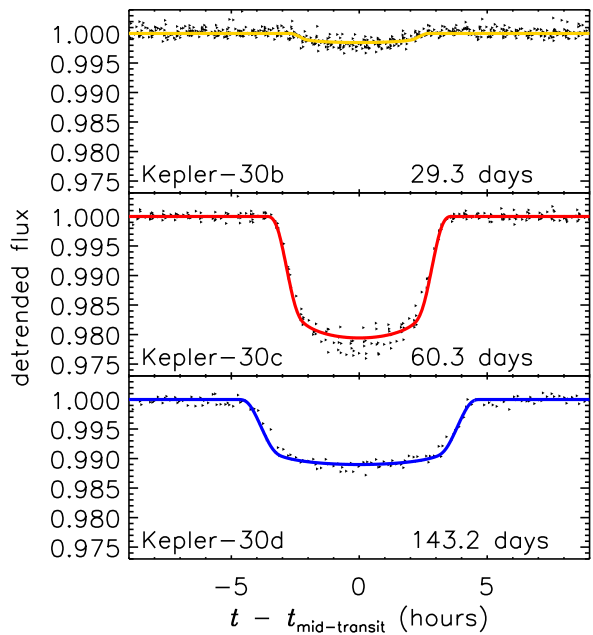}
\caption{\Kepb\ / KOI-806 phase curves, in the style of figure \ref{738phasecurve}.  \vspace{0.4 in}}
\label{806phasecurve}
\end{figure}

\begin{figure*}
\plotone{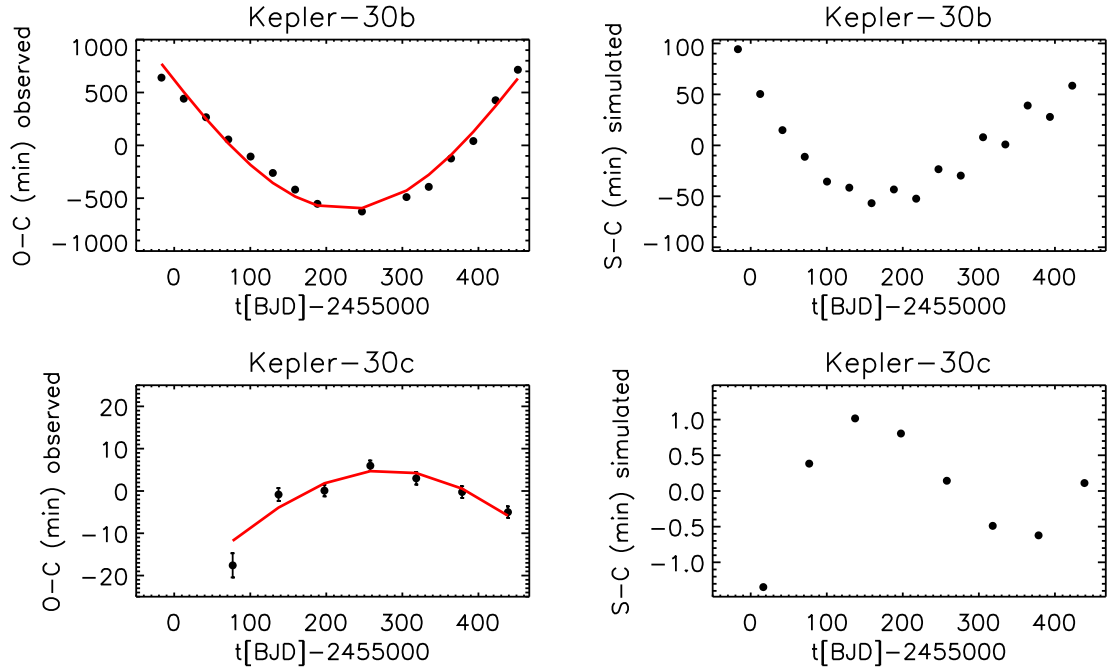}
\caption{\Kepb.  Transit timing diagram with theoretical model compared, for the interaction between \Kepb b and \Kepb c.  Panels are the same as in Figure~\ref{ot738}. \vspace{0.4 in}
}
\label{ot806_32}
\end{figure*}

\Kepb\ is a remarkable system of three planets, all with extremely significant transit timing variations.  Its lightcurve is shown in Figure~\ref{806lc} and each planet's phase curve is shown in Figure~\ref{806phasecurve}.  A Neptune-size inner planet ($P_b=29.2$~days) lies just interior to a 2:1 resonance with a Jupiter-size planet ($P_c=60.3$~days).  Their mutual perturbation causes a transit timing variation with a frequency equal to the difference of the orbital frequencies from the resonance:
\begin{equation}
P_{ttv} = 1 / | 2/ 60.3 d - 1 / 29.2 d | = 900 d. \label{eq:pttv806}
\end{equation}
Since one of the participating planets has deep transits and thus is presumably massive, the signal is expected to be large.  The theory predicts a sinusoid with a period a bit longer than the data.  Indeed, the data show an enormous signal consistent with these expectations (Fig.~[\ref{ot806_32}]).  Thus, we confirm \Kepb b and \Kepb c as objects orbiting the same star. 

Since their orbits are spaced a bit wider than the other planets confirmed in this paper, dynamical stability is not as strong a constraint on their masses.  We find a mass upper limit (see section~\ref{secMassLimits}) of about 17~$M_{\rm Jup}$.  However, eccentricity can also account for the large transit timing signals, so neither planet is necessarily nearly this massive (in particular, the small-radius \Kepb b should not be multiple Jupiter masses).  The deviations from a constant ephemeris for planet b are roughly parabolic and span $\pm 10$~hr, by far the largest timing variations for a multiple-planet system yet detected (but not quite rivaling the circumbinary planet Kepler-16(AB)b; \citealt{2011Doyle}).  However, there is more signal in this TTV curve than just the large parabolic trend.  The rising branch of both the observed signal and the theoretical signal show a short-timescale ``chopping'' effect for \Kepb b, where the even and odd transits times show structure, due to the 2:1 resonance.  A similar effect was seen previously in Kepler-9b \citep{2010Holman}. A mass limit that takes into account the observed transit times will be discussed in section \ref{sec:dynammodels}.  It results in a much tighter constraint, but it is less robust because it is not based on an exhaustive exploration of parameter space.  

\begin{figure*}
\plotone{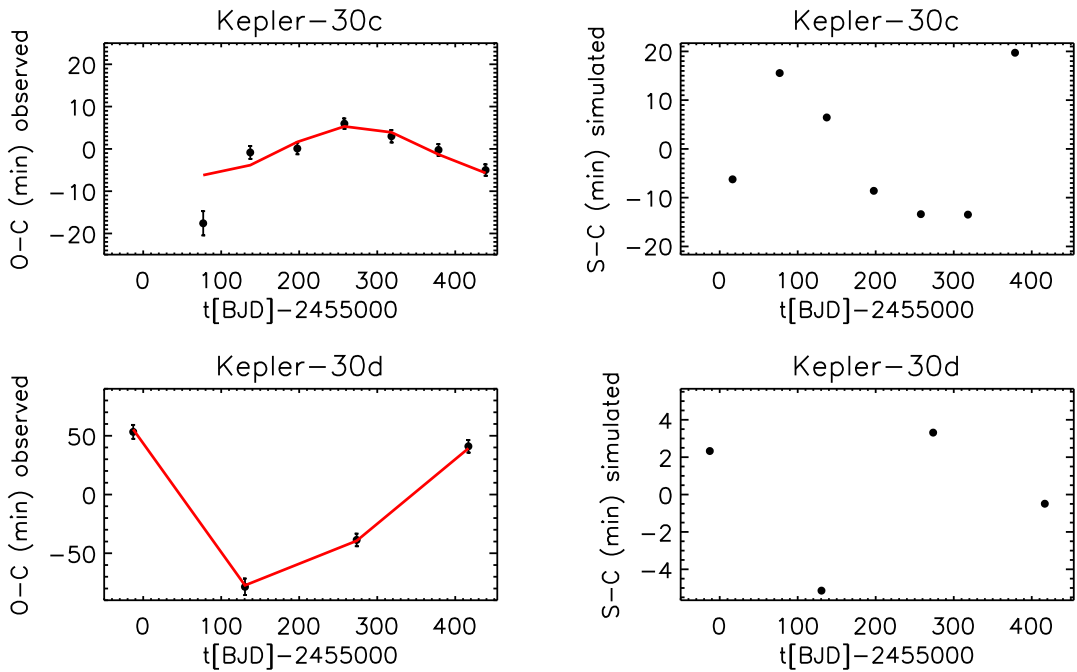}
\caption{\Kepb.  Transit timing diagram with theoretical model compared, for the interaction between \Kepb c (top) and \Kepb d (bottom).  (Ignore the labels above the plots for now, as they will be finalized with a Kepler number; these plots are for c and d.)  Panels are the same as in Figure~\ref{ot738}.\vspace{0.4 in}
}
\label{ot806_21}
\end{figure*}

We can also confirm \Kepb d is likely a planet in the same system, as transit timing variations with a $\sim2$~hr timescale are seen in it (fig.~\ref{ot806_21}).  However, the interpretation is slightly less clear, as only 4 transits of \Kepb d ($P=143.2128$~d) have been recorded so far, so the shape of the variation is poorly determined.   This perturbation is anticorrelated with the swing seen in its neighbor (\Kepb c), but that swing might be predominantly due to \Kepb c's near-resonant interaction with \Kepb b.   According to the nominal integration, the timescale for the interaction between planets d and c should be $\sim400$~days (Fig.~\ref{ot806_21} and Table~\ref{tabPlanetPairs}), similar to the span of the dataset.  When the dataset is doubled, this might be discerned from the longer timescale (eq.~\ref{eq:pttv806}) interaction between the inner two planets.  Even without a detailed analysis, we determine from transit timing that \Kepb d is an object in the same system.  We carry out preliminary transit-timing modeling of the three planets in section~\ref{sec:dynammodels}.

\begin{figure}
\plotone{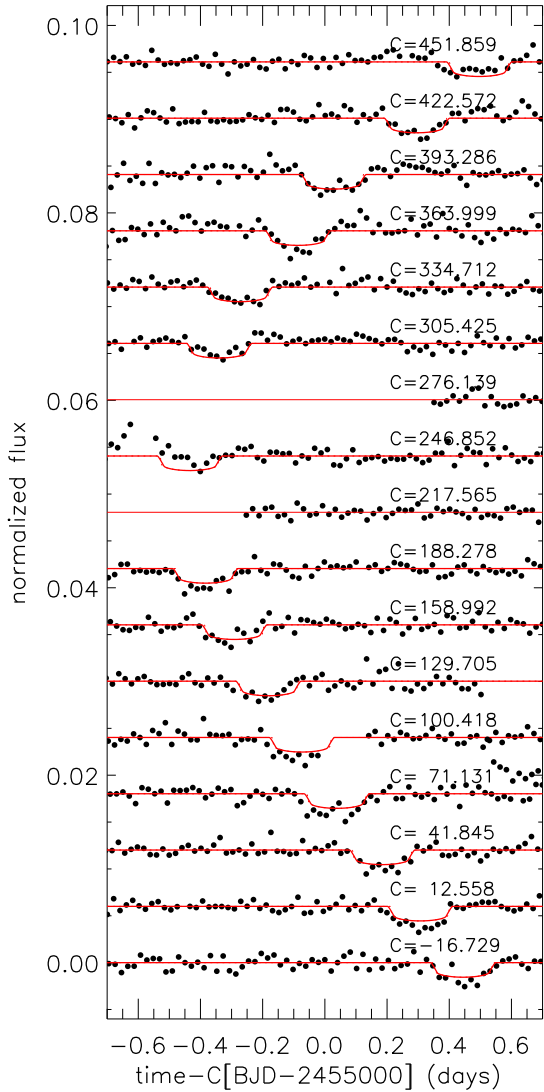}
\caption{ Individual transits of \Kepb b.  Each portion of the lightcurve is shown with its transit shifted to the best linear-ephemeris value, and the model is shown with the best-fit transit time taken into account.  This is the biggest transit timing variation yet reported in multiple-planet systems. \vspace{0.4 in}}
\label{80603_indiv}
\end{figure}

Due to the large timing variations and deep transits found in this system, we find it instructive to show individual transits to make several other points.   In Figure~\ref{80603_indiv} we plot the transits of \Kepb b.  Each portion of the lightcurve is centered where the transit would be if it followed the best-fit linear ephemeris.  Instead, we see that the timing of the planet varies by of order a day.  The standard \Kepler pipeline \citep{2010Jenkins} assumes near-periodic transits, but the transits are deep enough for detection despite its strong acceleration.  If the transits were considerably shallower, it may not have yielded a significant signal before the phase changed by more than a transit duration \citep{2011Garcia}.  To address this, the transit-timing subteam of \Kepler is currently developing search algorithms that assume either a quadratic, a sinusoidal, or a non-specified quasi-periodic ephemeris.  

\begin{figure}
\plotone{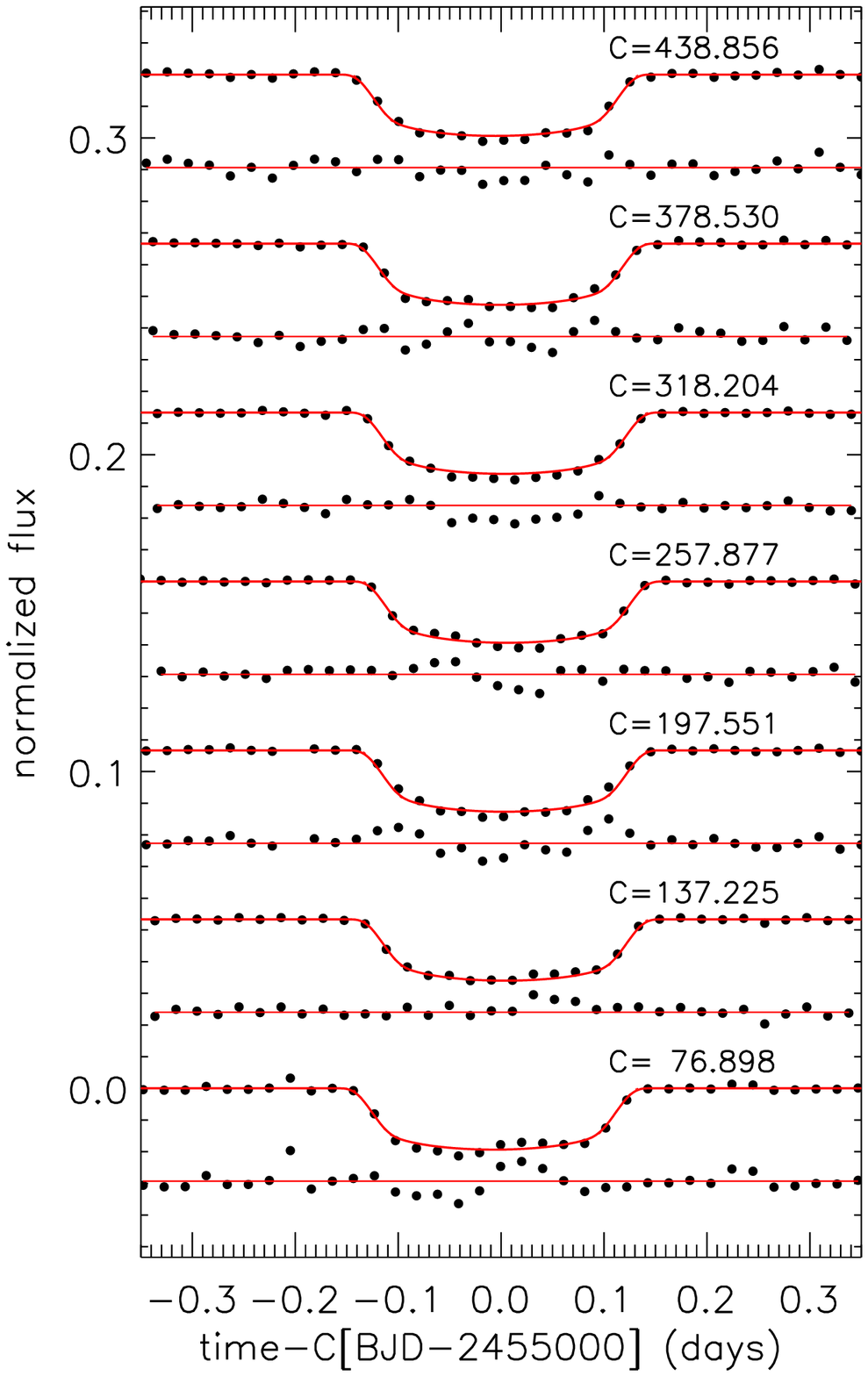}
\caption{ Individual transits of \Kepb c (KOI-806.02, the planet observed by \citealt{2011Tingley}), in the style of figure~\ref{80603_indiv}.  Below each transit is shown the residuals from the model, scaled up by a factor of 3 to better show the deviations.  The largest variations appear in eclipse and are correlated point-to-point.  Large starspots are visible in the lightcurve; we hypothesize this excess variability comes from the planet transiting over them.  \vspace{0.4 in} }
\label{80602_indiv}
\end{figure}

Finally, consider Figure~\ref{80602_indiv}, the individual transits of  \Kepb c.  The residuals (with a 3$\times$ zoom) are shown beneath each transit.  The average transit lightcurve does not capture some of the variations from transit-to-transit, at the level of $\sim 10\%$ of the planet's depth.  We suggest this is due to the planet transiting over starspots, which are clearly visible in the out-of-transit curve at the level of a few percent (fig.~\ref{806lc}).  To extract the spot period from the lightcurve, we used a slightly modified version of the Discrete Correlation Function of \cite{1988Edelson} (see also \citealt{1994White}), as was recently applied to the CoRoT-7 data \citep{2009Queloz}, and obtained $P_{\rm rot}=15.25 \pm 0.25$~days.  Modeling the spot-crossing signal during transits may allow the measurement of the spin orientation of the host star relative to the planets' orbits in this particularly valuable multiple-planet system \citep{2011SanchisOjeda,2011Nutzman,2011Desert}.   \Kepb c (also known as KOI-806.02) was recently detected in transit from the ground, by \cite{2011Tingley}, who made the point that it will serve as an interesting system to continue following even after \Kepler finishes its mission.

%\epsscale{0.8}

\epsscale{1.0}

%\epsscale{0.6}
\epsscale{1.0}

%\clearpage

\subsubsection{ KOI-935 = \Kepc}

\begin{figure*}
\plotone{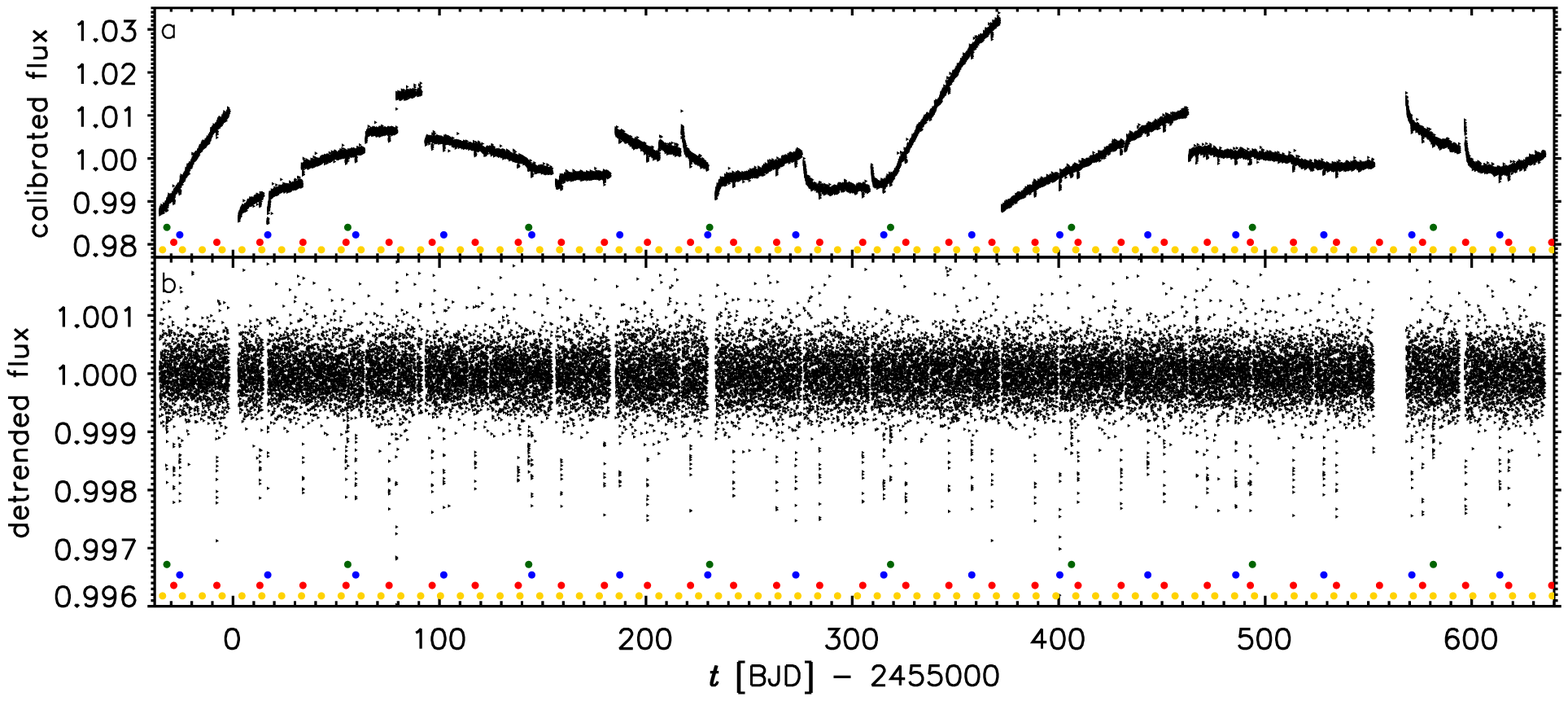}
\caption{\Kepc\ lightcurves, in the style of figure \ref{738lc}.  \vspace{0.4 in}}
\label{935lc}
\end{figure*}

\begin{figure}
\plotone{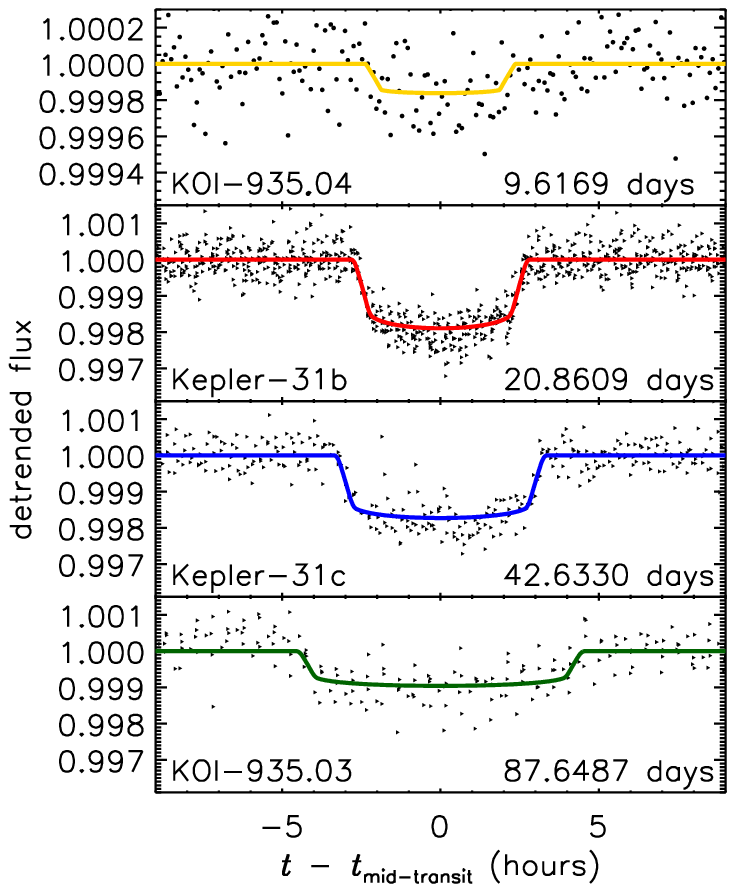}
\caption{\Kepc\ phase curves, in the style of figure \ref{738phasecurve}.   For the small inner candidate KOI-935.04, the phase is with respect to a linear ephemeris, the data in that panel are binned together in phase.  The vertical scale of that panel is 20\% of the other panels.  \vspace{0.4 in}}
\label{935phasecurve}
\end{figure}

\Kepc\ is a system of four planet candidates spaced just wide of a 1:2:4:8 resonant configuration.  Its lightcurve is shown in Figure~\ref{935lc} and its phase curves in Figure~\ref{935phasecurve}.  From the nominal models, only the middle two (\Kepc b and \Kepc c) are expected to show a $>1$~min variation, big enough to be detected in \Kepler data.  The timescale due to this near-resonance is predicted as 
\begin{equation}
P_{ttv} = 1 /  | 2/ 42.6330 d - 1 / 20.8609 d | =  980 d. 
\end{equation}

\begin{figure*}
\plotone{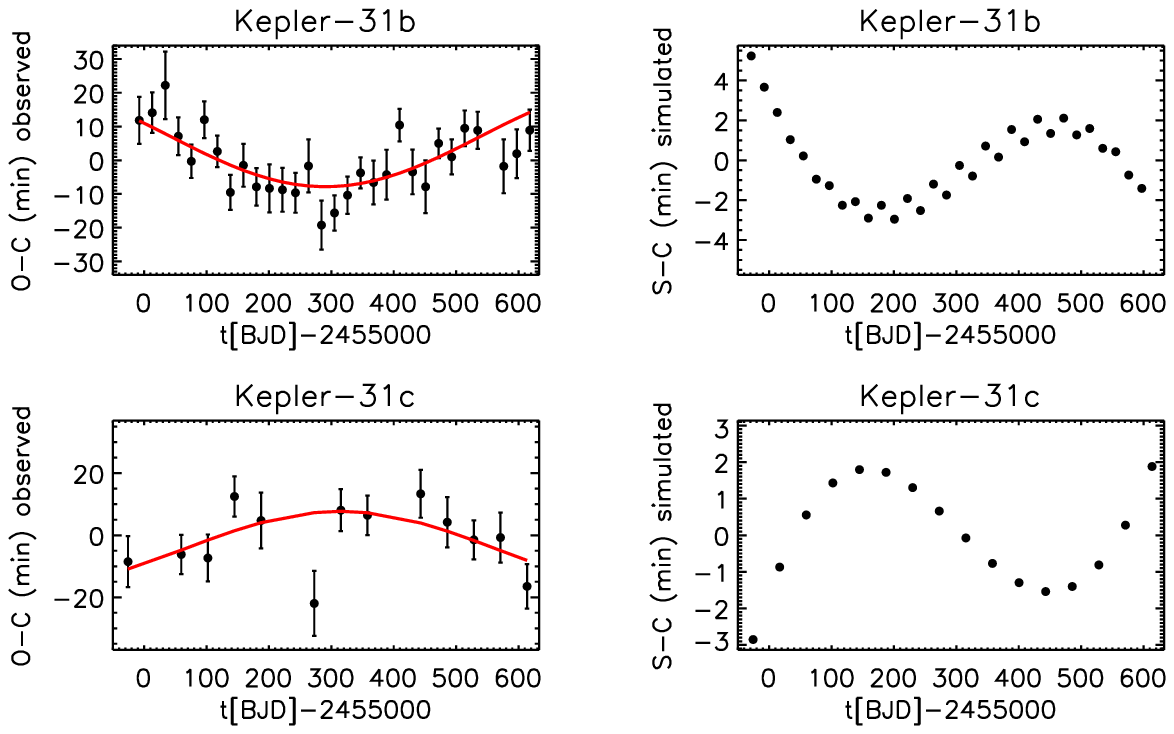}
\caption{\Kepc.  Transit timing diagram with theoretical model compared.   The data for planets b and c show  curvature on a long timescale, as predicted by a theoretical model in which they torque each other due to the close 2:1 resonance. 
\vspace{0.4 in} }
\label{ot935}
\end{figure*}

Thus, we have the now-familiar situation that a long-timescale variation is predicted due to the interaction.  In this case we supplemented the dataset with quarters 7 and 8 as well, which were able to capture more of this long-timescale variation.   The data for \Kepc b indeed clearly show a variation matching the prediction (Fig.~\ref{ot935}, top panels), with a low false alarm probability.  We are seeing its orbital period increase due to the torque exerted by a 2:1 mean-motion resonance.  The magnitude of this torque depends on both the planets' masses and eccentricities, so it is not surprising the theory underpredicted the amplitude by a factor of $\sim 3$.  Thus we confirm that these two objects are interacting, which combined with the mass limits set in section~\ref{secMassLimits}, confirms them as planets.  

The hint (FAP=$3.4\%$) of TTVs of \Kepc c due to \Kepc b does not itself satisfy our statistical criterion.  Nevertheless, the clear detection of TTVs in Kepler-31b consistent with the timescale predicted due to Kepler-31c provides evidence that they are in the same physical system. 

The system also has a long-period candidate KOI 935.03 and an inner candidate KOI 935.04, but these are not visibly interacting.  

%\epsscale{0.8}

%\epsscale{1.0}

%\clearpage

\subsubsection{ KOI-952 = \Kepd}

\begin{figure*}
\plotone{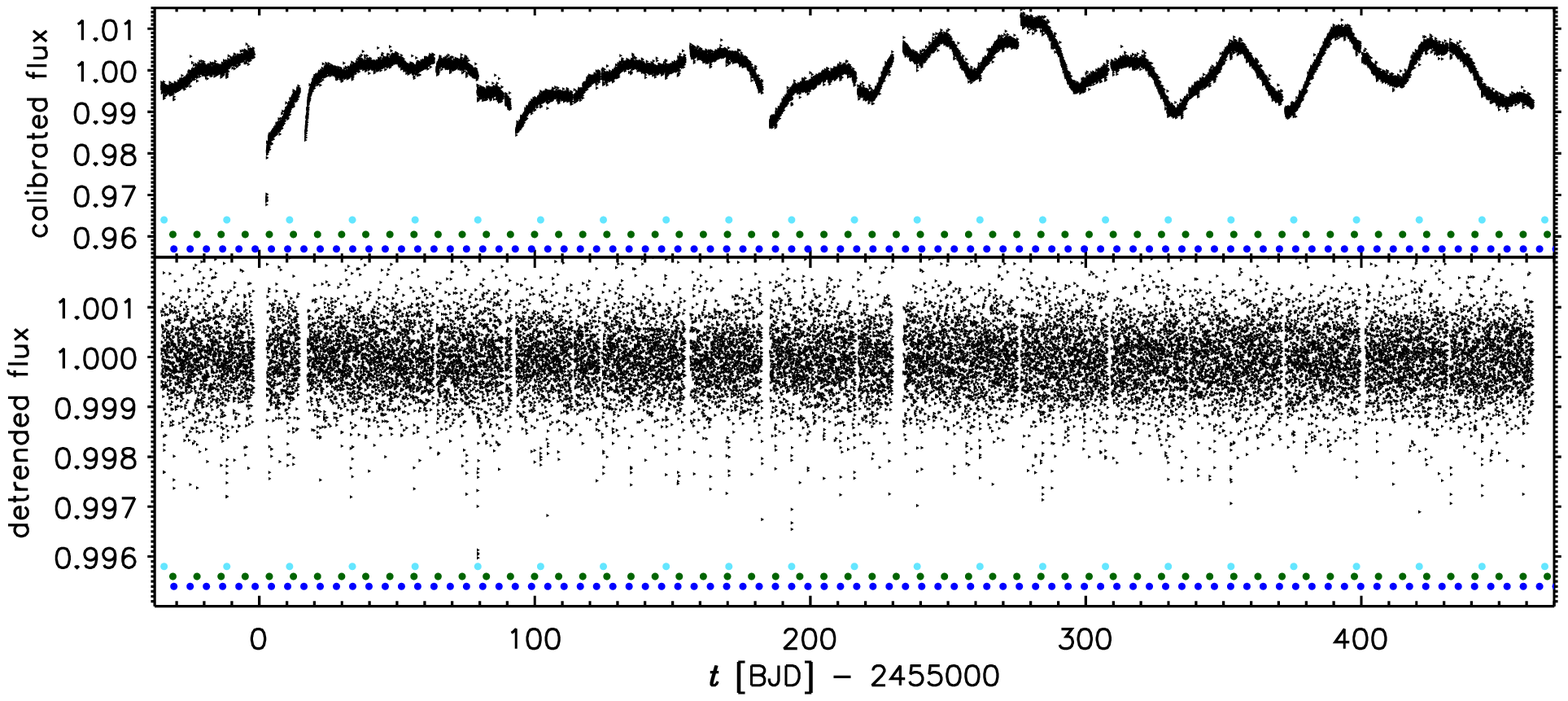}
\caption{ \Kepd\ lightcurve.   \emph{Upper panel:} quarter-normalized calibrated lightcurves; the stellar rotation period of  $37.8 \pm 1.2$~days is evident in quarters 5 and 6.   \emph{Bottom panel:} The bottom, middle and top rows of dots (blue, green, teal) are for \Kepd b, \Kepd c and KOI-952.03.  (The short-period candidates KOI-952.04 and KOI-952.05 are not detected in individual transits, so no pointers are given for them.) \vspace{0.4 in}}
\label{952lc}
\end{figure*}

\begin{figure}
\plotone{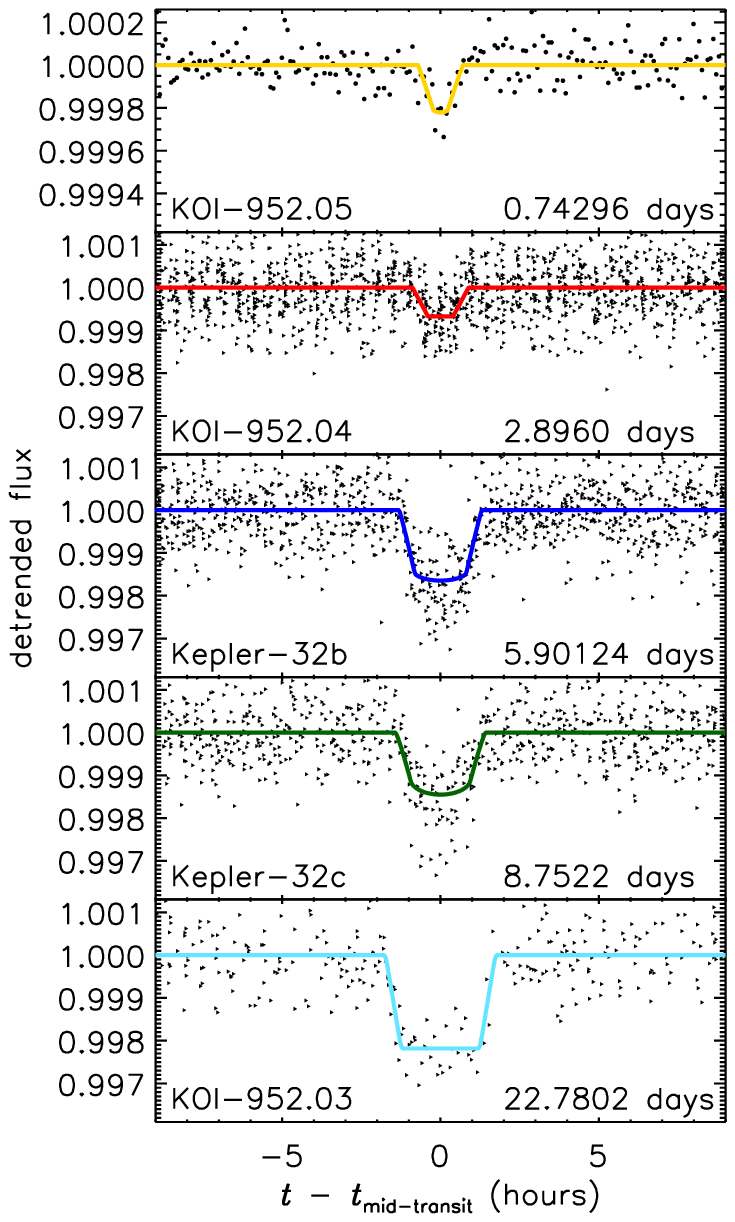}
\caption{\Kepc\ phase curves, in the style of figure \ref{738phasecurve}.  For the small inner candidate KOI-952.05, the phase is with respect to a linear ephemeris, the data in that panel are binned together in phase.  The vertical scale of that panel is 20\% of the other panels.  \vspace{0.4 in}}
\label{952phasecurve}
\end{figure}

\Kepd\ boasts five planet candidates, one of the richest systems among the \Kepler discoveries \citep{2011Lissauerb}.   Its lightcurve and phase curves are plotted in Figures~\ref{952lc} and \ref{952phasecurve}, respectively.  The lightcurve has variations suggesting a starspot signal, in which case we determine $P_{\rm rot}=37.8 \pm 1.2$~days using an autocorrelation technique as above. 

\begin{figure*}
\plotone{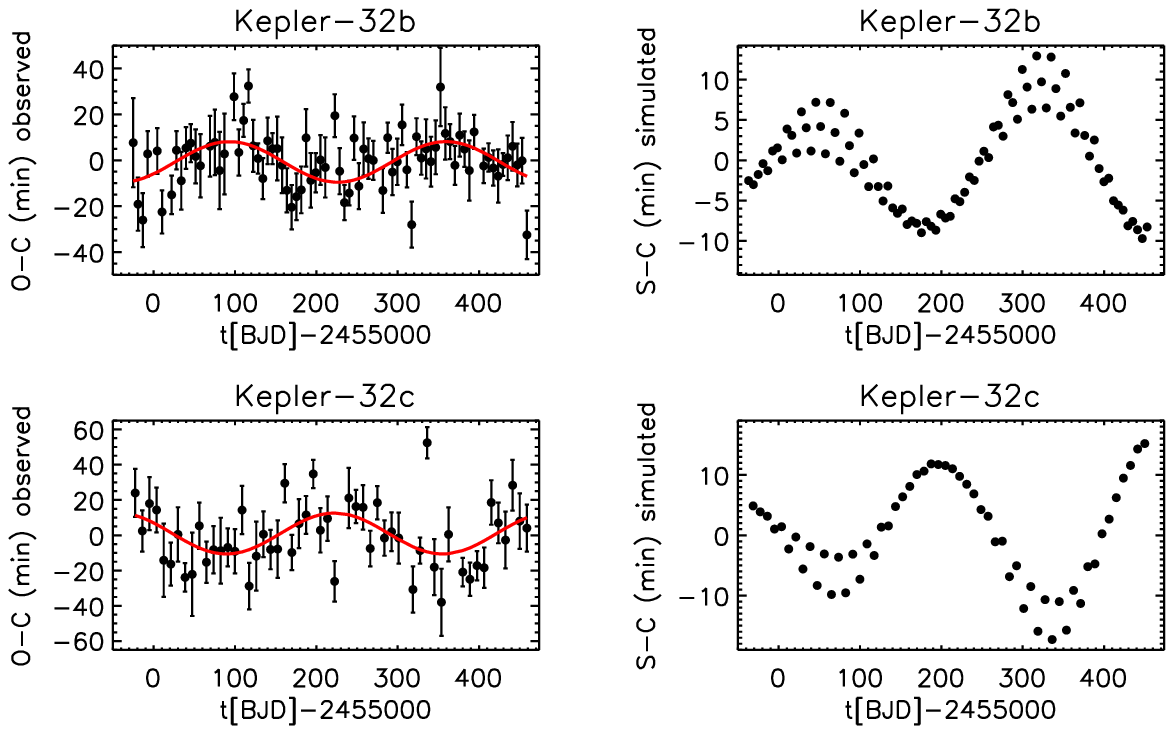}
\caption{Comparison of measured transit times (left) and transit times predicted by the nominal model (right) for a system containing only \Kepd b (top) and \Kepd c (bottom).  Details are described in the caption to Fig.\ \ref{ot738}. \vspace{0.4 in}}
\label{figTheory952}
\end{figure*}

Candidates .01 and .02 have short orbital periods and lie close to a 3:2 mean-motion resonance: $P_{.02}/P_{.01}=1.483$.  Therefore, we expect a short TTV timescale:
\begin{equation}
P_{ttv} = 1 / | 3/ 8.7522 d - 2 / 5.9012 d | = 260 d. \label{eq:ttv952}
\end{equation}
The observed timescale of TTVs is consistent with this expectation from the nominal N-body integrations (Fig.~\ref{figTheory952}).  The amplitude is somewhat larger than predicted by the nominal model ($\sim~2.6$ and 3.3 $\times$).  This is not unexpected as even modest eccentricities can significantly increase the amplitude and completely change the phase of the predicted TTVs.  The amplitudes also scale with the planet masses, which could be larger than those adopted in our nominal model.  The ratio of TTV amplitudes for \Kepd b and \Kepd c agrees with the predictions of the nominal model to $\sim~20\%$.  Despite these uncertainties, the requirement of dynamical stability provides firm upper limits on the planet masses.  Thus, we confirm this pair of planets and rename them as \Kepd b and \Kepd c.

Based on the nominal N-body integrations of the five planets, we expect shorter timescales and smaller amplitudes for transit variations of KOI-952.03, .04, and .05.  Indeed, no significant transit time variations were seen in them, nor do they induce detectable variation in \Kepd b or \Kepd c, so we cannot confirm them via TTVs at this time.  

%\clearpage

%\epsscale{1.0}

%\epsscale{0.7}

%\epsscale{1.0}

\subsection{ Identification of the Host Star}
\label{stellar_props}

In this section we discuss the probability that the planets are actually hosted by a star other than the target.  The area on the sky in which that other star is allowed is limited via lack of movement of the centroid during transits (\citealt{2011Torres}, Bryson et al. in prep).  Because of the low space density of transiting planetary systems, it is generally rather unlikely (quantified for each candidate below) that it is hosted by an unrelated background or foreground star, which is a minority contribution to the light of the target.  In comparison, because most stars are components of binaries, it is not inconceivable that our targets are physically bound binary stars where the second component is not detected, and the fainter component hosts the interacting planets (probability quantified below).   To pursue these calculations, we assume a maximum planetary radius consisting of the upper envelope of the mass-radius relation of known exoplanets.  If the radii implied by a blend scenario are either above that limit or imply masses too large for dynamical stability given the configuration of planets, then we reject that blend scenario.  Now we discuss details of each system. 

\subsubsection{\Kepa}

\begin{figure}
\plotone{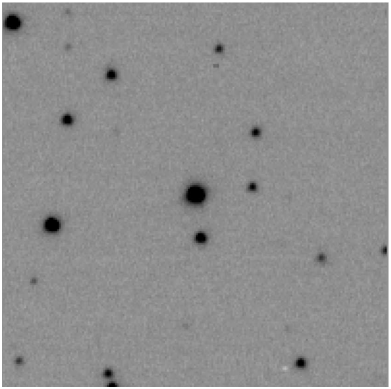}
\caption{ FTN optical (Sloan r) image of \Kepa.   The scale is 1 arcmin on each side, and north is up. }
\label{image738}
\end{figure}

\Kepa\ has low contamination from surrounding stars.  The nearest stars to \Kepa, listed in the \Kepler Input Catalog (KIC), are KID 10358756, a $Kp=17.5$~mag star located 6.5 arcsec to the south, and KID 10358742 located 9 arcsec away.  We examined the UKIRT J-band image and obtained a new image from Faulkes Telescope North in the Sloan-r band (figure~\ref{image738}); neither showed evidence of additional companions to a magnitude difference of $\sim 5$ in to $\sim 1$~arcsec.  We therefore adopt the aperture contamination calculated using stars in the KIC alone.

The planets pass the centroid test: the photocenter out-of-transit and during transit are consistent to within ($2 \sigma$, $1 \sigma$) for planets (.01,.02) respectively.  This rules out out background stars as the planet host if they are further than $Rc = (0.7, 0.8)$~arcsec from the target, where $Rc$ is the 3-$\sigma$ radius of confusion.  Thus none of the background stars in the image can be the host.   Next we computed the space density of background stars, to determine the chance that the planetary system is actually around a star other than the target.  That is, we assume that the TTV signature robustly indicates this system is planetary, but we wish to see the probability it is around a star other than the one we have characterized. 

Stability limits (section~\ref{secMassLimits}) require that the planets cannot be too massive relative to their host, otherwise they would be unstable.  In this case, the mass ratio of both planets to their host must be $\lesssim 2 \times 10^{-4}$.  With such low masses, we consider $1 R_{\rm Jup}$ to be a robust upper limit for the radii of the planets.  This sets the constraint on the maximum magnitude difference that the blending star can have. Any star fainter than this limit would not contribute enough to the combined flux to generate an eclipse as deep as observed (830 ppm).  Based on the U-shaped light-curve, we assumed that grazing transits were excluded, such that $(R_p/R_\star)^2$ is a good approximation of the (undiluted) depth.  We used Yonsei-Yale isochrones (3 Gyr, solar metallicity) to relate the maximum allowed magnitude difference ($\Delta$~Kp) to the spectral type and mass of the blending object. 

The background host hypothesis splits into two distinct cases: the planets are larger and (1) orbit a physically unassociated star, or (2) orbit an unseen binary companion of the target.   

For the former, we used the Besancon model of the galaxy \citep{2003Robin} to estimate the number of background stars within the region allowed by the centroid shift result {\it and} the region in the allowed magnitude difference (or stellar mass).  The probability of having a background star in this region is 0.6\%.  If we assume (conservatively) that the more massive planets on the background star are as likely, a priori, as less massive planets on the target star, then the odds ratio that the target star hosts the planets rather than a background star is $\sim 170:1$.  

Next we calculate the probability of the latter case, that the planetary system is orbiting an unseen stellar companion to the target star.   The blends involving stars less massive than a $0.6 M_\odot$ star would need large planets, which in the range of known exoplanets would be too massive to be stable.  In the range $0.6$-$0.8M_\odot$, we can use the (Sloan) r - (2 MASS) K color of the target to rule out such blends, as they would not correspond to the observed color.  A blended star more massive than $0.8M_\odot$ is allowed, and we call this scenario a `twin star' blend.   It is possible to put limits on such blends because we do not see two stars in the images, nor do we see two sets of lines in the spectrum.  The only remaining case is that the two twins have a large separation along the line of sight (a small velocity difference) but not in the plane of the sky.   If there were such a case, it would correspond to twice the light, so the undiluted depth of transits would be twice as large, so the planets would be $\sim \sqrt{2}$ larger.  

Even without employing that constraint, we use \cite{Raghavanetal.2010, DuquennoyMayor91} to determine the distribution (frequency/mass ratio) of binaries and find the probability of having an unseen companion star in the allowed mass range ($0.8$-$1.0M_\odot$) is 7\%. 

For this target, which has a clean centroid and closely-packed planets that would go unstable if they were gas-giant masses, we find that background hosts are much less likely than the target star being the host. 

\subsubsection{\Kepb}

We obtained an I-band image of \Kepb\ from Lick 1-m telescope, and found no stars besides those in the KIC, the closest of which is $Kp=19.8$~mag KIC 3832477, 7.6 arcsec to the ENE.  A star $\sim 3$~arcsec to the E is marginally detected in J-band with UKIRT, suggesting it is $\gtrsim 5$ magnitudes fainter than the target in Kp.  It cannot host the $2\%$ deep transit of \Kepd c, and we neglect its additional contribution to the dilution.  Furthermore, since planets c and d produce deep transits, their centroid information is particularly valuable.  Their transits have a consistent centroid with the target to $\sim1 \sigma$ in either case, and they limit the distance of a putative blend hosting the system to be within $Rc=0.2$~arcsec of the target.

As a routine part of vetting planet candidates, the depth for odd numbered transits was compared to the depth for even numbered transits.   For the other candidates reported in this paper, there is agreement to $\sim 3 \sigma$; however in this system, they disagree by $16\sigma$ for c and $4\sigma$ for d.  In some cases, that would point to a near-twin blended eclipsing binary causing the events.   In this case, we have already identified via figure~\ref{80602_indiv} that transiting over starspots is the probable cause of these depth variations. 

Using the same procedure as for \Kepa, we find the probability that an unassociated background star hosts the planets is $\sim2 \times 10^{-4}$.  In this case, a physical binary companion cannot host the planets, as then their depths would be large despite dilution, and their inferred radii would be larger than any planetary radii thus far measured.  

\subsubsection{\Kepc}

\Kepc\ has low contamination from surrounding stars; no stars are seen within 8 arcsec in a UKIRT J-band image, and the closest comparably bright star is KIC 9347893, 9.4 arcsec to the west.  Moreover, the centroid information has all transits coincident within $1 \sigma$ of the target.  The transits cannot be hosted by a background star further than $Rc=(0.3, 0.5, 0.8)$ arcsec in the case of \Kepc b, \Kepc c, KOI-935.03 respectively.   For KOI-935.04, the transits are too shallow for a constraining centroid analysis. 

Again pursuing probability calculations as above, the chance of a star unassociated with the target being the actual host is only $\sim3\times10^{-4}$.  The probability of a physical companion hosting the planets is $\sim 0.04$.  

\subsubsection{\Kepd}

A J-band image from UKIRT shows the nearest star to be KID 9787232, $\sim~6.6"$ to the west, resulting in rather low contamination. 

The centroids during transit for \Kepd b and \Kepd c differ from those out-of-transit by only $\sim~2\sigma$, roughly consistent with measurement uncertainties.  The $\sim~3\sigma$ radii of confusion $Rc$ are 0.5" for \Kepd b and 0.8" for \Kepd c.  For KOI-952.03, .04, and .05, the transits are too shallow for a constraining centroid analysis. 

The host star is an M-dwarf and therefore of special interest.  The \Kepler Follow-up Program has obtained two spectra of \Kepd: one spectrum from McDonald Observatory 
%(HJD=2455468.599753) 
and one from Keck Observatory.
%(HJD=)
Both spectra are weak due to the faintness of the star (Kp=15.8).  
The cross correlation function between the observed spectra and available models is maximized for temperatures of $\sim3900$~K and $\sim3600$~K, respectively.  However, the atmospheric parameters are not well determined, as the star is cooler than the library of atmosphere models available.  Both spectra are consistent with the KIC classification as a cool dwarf  ($T_{\rm eff}=3911$, $\log g=4.64$, [M/H]=0.172). We conservatively adopt these values of $T_{\rm eff}$ and $\log g$ with uncertainties of 200K and 0.3 dex and a [M/H] of $0\pm~0.4$ based on the KIC \citep{2011Brown}.  By comparing to the Yonsei-Yale isochrones, we derive values for the stellar mass ($0.58\pm~0.05 M_\odot$) and radius ($0.53\pm~0.04 R_\odot$) that are slightly larger than those from the KIC.  We estimate a luminosity of $0.06\pm~0.02~L_\odot$ and an age of $\le~9$Gyr.  

\cite{2011Muirhead} have also obtained high-resolution IR spectrum of \Kepd=KOI-952, finding a stellar $T_{\rm eff}=3726^{+  73}_{ -67}$, [Fe/H]$=0.04^{+0.08}_{-0.10}$.  Interpreting their data via Padova models \citep{2002Girardi}, they inferred a considerably less massive and smaller star.  We encourage further detailed analyses of the host star properties, as these have considerable uncertainties that directly affect the sizes and masses for the planets.  

The probability of a star unassociated with the target being the actual host is only $\sim3\times10^{-3}$.  The probability of a physical companion hosting the planets is $\sim 0.34$.  This latter number is relatively large in this case because all the transit depths are small, so they could in principle be much larger planets hosted by a star which is dramatically diluted.  This opens up the possibilities for a very large range of companions (down to masses as low as $\sim0.1 M_\odot$) that could host one or more of these objects, as long as transits near apocenter are invoked to match the durations (fig.~\ref{durations}).

\epsscale{0.7}

%\clearpage 

\section{ Planetary Mass Limits }
\label{sec:Masses}

\subsection{Dynamical Stability Analysis }
\label{secMassLimits}
Many of the systems in this paper and its companions (Papers II and III) are not completely solvable with present data; e.g., the gravitational interactions of the component planets do not yield unique solutions for their masses.  Rather, there exists degeneracy between the masses and eccentricities, as was the case for Kepler-9 before radial velocity constraints were applied \citep{2010Holman}.  However, we constrain them to be in the planetary regime because the pairs of planets all have small period ratios.  In two-planet systems, a sharp boundary exists between provably stable orbits \citep{1982MarchalBozis} and orbits that are allowed to cross, according to energy and angular momentum conservation.  This boundary is when the separation of the planetary semi-major axes, $a_{out}-a_{in}$, exceeds a certain number ($2\sqrt{3} \approx 3.46$, for coplanar, circular orbits) of mutual Hill spheres, 
\begin{equation}
r_H = \frac{a_{in}+a_{out}}{2}  \Big( \frac{ M_{in} + M_{out} }{3 M_\star} \Big)^{1/3}.
\end{equation}
When the separation is only slightly closer than this, numerical integrations generally show the planets chaotically interact and have close encounters after only several million orbits, an astrophysically short amount of time \citep{Gladman:1993,2006Barnes}.  An exception is that planets locked in mean motion resonances may avoid each other due to the correlated phases of their orbits; Neptune and Pluto are a familiar example \citep{1965Cohen}.  

We ran a suite of numerical integrations for each planetary pair with a significant TTV interaction to determine when the masses would be too large to yield a long-term stable system.  We used the HYBRID integrator within the \emph{Mercury} package \citep{Chambers:1999}.  Since we stopped after close encounters, the mixed-variable symplectic algorithm was exclusively used \citep{Wisdom:1991}, with a timestep 1/20th of the inner planet's orbital period.  An implementation of general relativistic precession was included \citep{1986Nobili,2011Lissauerb}.   We picked the planets' ratio of masses from their ratio of radii according to the relation $M_{in}/M_{out} = (R_{in} / R_{out})^{2.06}$ \citep{2011Lissauerb}.  We performed 6 integrations with differing total mass, starting at the theoretical stability boundary and becoming more massive by logarithmic steps of $0.25$~dex.  The orbits were taken as initially circular and coplanar, with phases determined from the \Kepler data.  

\begin{figure}
\plotone{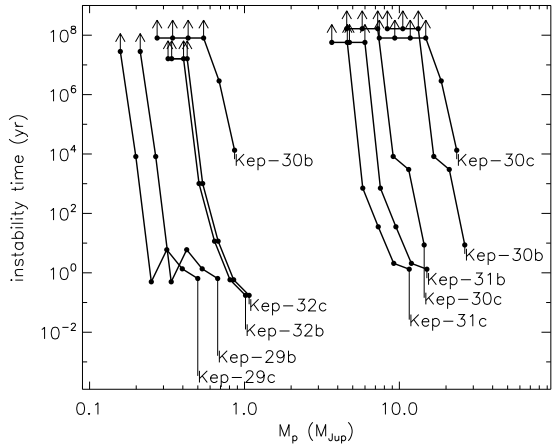} 
\caption{The time until instability, as a function of planetary mass.  This figure shows that for these objects to be stable, they must have masses traditionally associated with the planetary domain.}
\label{figMassLimits}
\end{figure}

We stopped the integration when planets came within 3~$r_H$\ of each other, calling this time the instability timescale.  We ran only one integration at each mass, so we recognize that this investigation only identifies the order of magnitude of that timescale -- however, consistent with previous investigators \citep{1996Chambers,1997Duncan} we find that a factor-of-two change in mass causes many orders of magnitude of difference in instability time.  Figure~\ref{figMassLimits} gives the instability timescales as a function of planetary mass.  We determine the mass upper limits based on the minimum mass for which the integrations start going unstable on $\lesssim 10$~Myr timescales; for this purpose, we do not need to run the integrations for the likely lifetimes of these systems.  These limits are given in Table~\ref{tabPlanets}.  These are quite conservative upper limits, and the true masses could all be quite smaller.  Full modeling of the systems, including eccentricities, is required for a true mass measurement via transit timing variations (e.g. Kepler-11 by Lissauer et al.).  Nevertheless, this exercise shows that all of the systems we are presenting should be considered ``planetary'' systems, rather than stellar systems. 

In this investigation, we have simulated only pairs of planets and neglected additional planets, either other transiting candidates or completely unknown planets.  When third planets are added, stability constraints have been found by numerical integrations to become even more stringent \citep{1996Chambers, 2008Chatterjee, 2010FM}.  An interesting exception was discussed by \citealt{2005RaymondBarnes}, but the third planet in that case quenched secular eccentricity cycles, which is not the chaotic mechanism for instability we have investigated above.  Therefore the future confirmation or discovery of such planets will not compromise the conclusions here. 

Similarly, we neglected possible eccentricity in the planets.  For non-resonant cases, non-zero eccentricities would only serve to bring the planets closer together at conjunctions, making them more unstable \citep{1989Duncan, 2007Zhou}.  However, eccentricities can actually increase the stability of resonant pairs.  This effect could prolong the lifetime of \Kepa, such that its true upper-limit masses are underestimated by at least a factor of several.  Since these orbits are closely packed however, resonances besides the 9:7 can overlap with it, leading to chaotic instabilities.  Using the $\Delta a / a > 1.3 \mu^{2/7}$ criterion for stability \citep{1980Wisdom}, the maximum planet-to-star mass ratio is $\mu=0.02$.  If that mass is shared among the planets, then they both fall in the planetary regime ($\lesssim 10 M_{\rm Jup}$).

\epsscale{1.1}

\subsection{ Preliminary Dynamical Fits to the Transits Times}
\label{sec:dynammodels}
We have used the method first developed for the Kepler-9 and Kepler-11 discovery papers \citep{2010Holman,2011Lissauera} to fit preliminary dynamical models to all the planetary pairs of this paper and the two companion papers.  

Using the Levenberg-Marquardt algorithm to drive 3-body numerical integrations, minimizing the $\chi^2$ of the residuals of the data minus the model.  The free parameters are the mass $M_p$ of each planet and the Jacobian coordinates at dynamical epoch BJD 2455220.0 of each planet: orbital period $P$, epoch of mid-transit $T_0$, and the in-sky-plane and perpendicular-to-sky-plane components of the eccentricity vector, $e \cos \omega$ and $e \sin \omega$, respectively.   We ignore the dynamical effect of any planets that remain candidates in this work, just focusing on the interaction between the planets that show significant TTVs.  

The resulting fit parameters may be found in Table~\ref{tabSolutions}.  In many cases, the eccentricities and masses are very highly correlated, resulting in poor errors on each quantity.  However, we report many more decimal places beyond what seems significant, as to make these fits reproducible by others.  We note the contribution to $\chi^2$ for each planet and the number of degrees of freedom (d.o.f.) for that planet, meaning its number of data points minus the 5 free parameters used to model it.  (In two-planet fits, there were 10 free parameters total.)  In fits that are statistically acceptable, $\chi^2$ should equal d.o.f. to within a few times $\sqrt{{\rm d.o.f.}}$.  Of the systems confirmed in this paper, one of the poorer fits was for \Kepb\ / KOI-806, which might be attributable to starspots' effect on estimates of transit times. 

We found that \Kepa, the system at the 9:7 resonance, needed special treatment.  Allowed to fit freely, the eccentricities solved for large values, causing the planets to cross.  Because of phase protection by the resonance, this was allowed by the data, but after a secular timescale the planets began chaotic scattering.  Therefore in the fit reported in Table 5, we restricted the absolute value of each eccentricity components to be less than 0.05.  The result was integrated using the Bulirsch-Stoer algorithm in \emph{Mercury} \citep{Chambers:1999} for 30~Myr, during which it showed stable and regular orbital oscillations.  In other systems the eccentricities are quite moderate compared with the separation in semi-major axis, and we have not verified stability for these.  In future detailed fits to the data, stability could be used as a principle guiding the results. 

In only a few cases are the masses meaningfully measured, according to the formal error bars (e.g. $\gtrsim 3 \sigma$).  These error bars sample only the local minimum of the fit.  We recognize that drawing meaning from the local curvature about the entire probability surface is hazardous.   However, we hope this preliminary work on transit fits will inspire other investigators to exhaustively explore orbital configurations that fit the data.  In the meantime, we tried to set $3 \sigma$ upper limits on the masses of the planets via the method developed in \cite{2010Holman}: we moved the mass of one of the planets away from the best-fit, incrementing by $\sim 50\%$, and solving for the other parameters each time.  When $\chi^2$ had grown to $9$ greater than the $\chi^2$ of the best-fit, we used the mass of the planet in that fit to define the $3 \sigma$ upper limit.  These limits are reported in table~\ref{tabSolutions}.  In comparison to the stability study, these TTV limits on mass are much tighter.  However, this method is rather delicate, in that we have not fully explored the global parameter space for a possibly more massive planet.  The stability study provides more robust mass upper limits on the masses of these planets.

\section{ Discussion }
\label{sec:discuss}

%\subsection{Kepler}
%
% Kepler mission
%

Taking stock of the results, we have 
\begin{itemize}
\item developed a new approach to confirming planets, by testing whether the transit timing signal is consistent with that produced by a known perturber, a second transiting planet,
\item applied this approach to \Kepler transit timing data, and confirmed 9 planets in 4 planetary systems (and reconfirmed 16 planets in 8 additional systems), 
\item showed that their masses must be in the planetary regime via stability arguments, and provided dynamical fits to the data. 
\end{itemize}

The systems discussed herein have remarkable properties.  

\Kepa\ is clearly engaged in a second-order resonance, which has only been observed for the 3:1 resonance previously (HD 60532; \citealt{2008Desort,2009Laskar}).  In such cases, adiabatic capture from low-eccentricity is possible, but at a finite speed of migration, the capture probability is enhanced if planets migrated towards each other on eccentric orbits \citep{2010Rein,2011Mustill}.  Previously for systems of super-Earths, theorists \citep{2007Terquem,2010McNeil,2010Ida,2011Liu,2011Pierens} have suggested planets could be about this close to one another and trapped in resonances, but they have always focused on first-order resonances.  Within the context of those formation models, the relative proportion of higher order resonances is now a pertinent question. 

\Kepb\ is perhaps the most dramatic system described here, having properties similar to multiple planet systems long-known from Doppler surveys.  The inner planet b has a $>1$~day swing in its transit times, perturbed by its near-2:1 outer companion, c.  This situation is very similar to that predicted by \cite{2005A} for the 2:1-resonant pair GJ 876 b/c, as the outer two planets clearly have gas-giant size.  Continued monitoring of the transits may be able to give unique parameters for this system, and future information about duration variations or lack thereof may allow a measurement of their mutual orbital inclination.  Planet c has a clear spot-crossing signal; its depth varies as a function of the spot phase of its host star.  With careful spot modeling, this may allow the interpretation of the orientation of the host star's spin relative to the planets' orbits.  Moreover, its transits are rather deep, from which we infer a large radius (Table~\ref{tabPlanets}: $1.27 \pm 0.16 R_{\rm Jup}$).  This is larger than expected for a planet receiving so little irradiation: the theoretical models by \cite{2007Fortney} are typically smaller than $\sim1.2 R_{\rm Jup}$, and \cite{2011Demory} confirmed that observationally using \Kepler giant-planet candidates.  We entertained the notion that this increased depth might be due to rings \citep{2011Schlicting}, but unfortunately the characteristic ``shoulders'' were not seen in the transit lightcurve. 

\Kepc\ has a 20.9-day planet and a 42.6-day planet now confirmed.  Two other candidates, at 9 and 88 days, are present yet remain unconfirmed.  If they can be confirmed at a later date, this system will be in an extraordinary near-1:2:4:8 resonant chain.  Evidence for that sort of architecture has been building on several different fronts: radial velocity detections (HD 40307: \citealt{2009Mayor}; GJ 876: \citealt{2010Rivera}; HD 20794: \citealt{2011Pepe}), and a dramatic 4-planet system discovered by direct imaging (HR 8799: \citealt{FabryckyMurray-Clay2010, Maroisetal.2010}).  It has even been suggested that the Solar System giant planets began in a multi-resonant configuration \citep{2007Morbidelli, 2008Thommes}, albeit with links more compact than 2:1 resonances \citep{2001Masset,2008Pierens}. 

\Kepd\ is a particularly clean case of anticorrelated transit times that also have a timescale matching baseline expectations.  It is particularly interesting to confirm these planets, because their host star is an M-dwarf, around which small planets appear to be particularly abundant \citep{2011Howard}.  The lightcurve boasts 3 additional candidate planets as well, making a particularly rich system. 

We can attribute most of these signals to a slight displacement of the planets' mean motions from a strict commensurability: they are involved in a ``great inequality,'' the orbital element oscillations of Jupiter and Saturn due to their displacement from the 5:2 resonance.  The same sort of perturbations are detected in the system of planets orbiting PSR1257+12 \citep{1992Rasio,1992Malhotra,1993Peale,1994Wolsz}.  The planets torque each other's orbits, resulting in a period increase or decrease, depending on where the location of conjunctions is relative to their periapses and to the line-of-sight.   The departure from commensurability causes this location to move, generating a predictable fluctuation in their orbital periods.  The timescale for this fluctuation is easily calculated from their orbital periods: equations \ref{eq:ttv738}-\ref{eq:ttv952}.  A general feature is that large-amplitude timing signals take many orbits to manifest themselves.  If the \Kepler mission is extended to eight years, this method would reach its full potential for planetary pairs with longer ``great inequality'' or libration timescales (e.g., \Kepa, fig.~\ref{ot738}).  As the \Kepler mission seeks to confirm longer-period candidates, in particular candidates in the habitable zone, we will be attempting transit timing analyses based on fewer transits.  Dynamical theory may be required to condition our expectations about transit timing variations.  This paper is a stepping stone to that type of analysis.  

\begin{figure}
\plotone{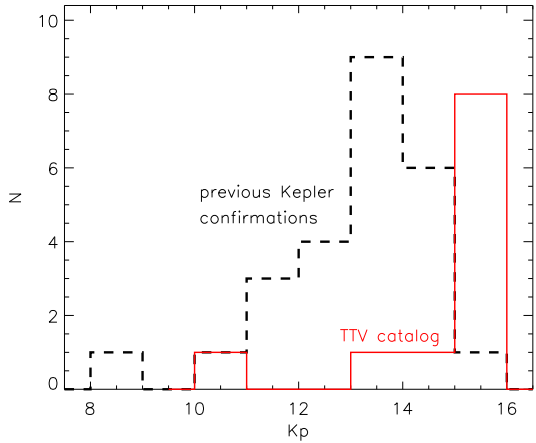}
\caption{ Histogram of stellar magnitudes in the \Kepler band (Kp).  The TTV catalog of planets confirmed by their interaction with each other (the $10$ planetary systems from Paper II, Paper III, and this paper) is not dependent on follow-up with other telescopes, so it more uniformly samples the intrinsic magnitude distribution of \Kepler stars.  The previously confirmed planets, by the \Kepler team (Kepler-4-Kepler-22) and also by \cite{2011Santernea, 2011Santerneb,2011Bouchy,2011Bonomo,2011Johnson}, do depend on other telescopes, and thus tend towards brighter targets.  \vspace{0.4 in}}
\label{fig:maghist}
\end{figure}

Along with Papers II and Paper III, we have confirmed 21 planets in 10 systems.  The transit timing method has very little bias with respect to the magnitude of the stellar host, as planetary systems with large, detectable perturbations are hosted by stars throughout the sample.  This contrasts with other \Kepler confirmations so far, which have mostly relied on ground-based (and \emph{Spitzer Space Telescope}) follow-up to confirm the planetary nature of the transiting objects.  In particular, we show in Figure~\ref{fig:maghist} that the current TTV confirmations are drawn from a very much fainter population than the previous confirmations. 

In two of these systems, as well as systems in Papers II and III, additional candidate planets have been found, but not confirmed via TTV or other methods.  \emph{A priori}, it has been argued that planet candidates in multiple-planet systems have a higher fidelity than planet candidates that are by themselves, even before performing follow-up observations or analysis (\citealt{2011Lissauerb,2011Latham}, Lissauer et al. in press).   Previously our team has pursued further analysis to validate additional small planets in TTV-active planetary systems: Kepler-9d \citep{2011Torres}, Kepler-10c \citep{2011Fressin}, and Kepler-18b \citep{2011Cochran}.  For the candidates listed here, many have good limits on $Rc$, the distance to which an unseen blend is a possible source of these transits.  Their phased lightcurves are easily fit by planetary lightcurves (``U'' shaped) rather than preferring blended binary lightcurves (which are {\it usually} ``V'' shaped).  Finally, the additional candidates have durations that can be explained by orbiting the same stars as the confirmed planets (Fig.~\ref{durations}).  Despite the fainter target stars which makes formal validation difficult, we expect most if not all of these candidates are indeed planets.

\acknowledgements  Funding for this mission is provided by NASA's Science Mission Directorate.  We thank the entire \Kepler team for the many years of work that is proving so successful.  We thank E. Agol for comments and G. Sokol for assistance analyzing starspot variations. 
D. C. F. and J. A. C. acknowledge support for this work was provided by NASA through Hubble Fellowship grants \#HF-51272.01-A and \#HF-51267.01-A awarded by the Space Telescope Science Institute, which is operated by the Association of Universities for Research in Astronomy, Inc., for NASA, under contract NAS 5-26555.
E.B.F acknowledges support by the National Aeronautics and Space Administration under grant NNX08AR04G issued through the Kepler Participating Scientist Program.  This material is based upon work supported by the National Science Foundation under Grant No. 0707203.
This paper uses observations obtained with facilities of the Las Cumbres Observatory Global Telescope.

{\it Facilities:} \facility{Kepler}.

\clearpage

\begin{deluxetable}{lcccccccccc} 
\tabletypesize{\scriptsize}
\tablecaption{Properties of target stars}
\tablewidth{0pt}
\tablehead{
\colhead{} & 
\colhead{KOI} & 
\colhead{KIC-ID} & 
\colhead{Kp} & 
\colhead{C0\tablenotemark{a}} &
\colhead{C1\tablenotemark{a}} &
\colhead{C2\tablenotemark{a}} &
\colhead{C3\tablenotemark{a}} &
\colhead{CDPP\tablenotemark{b}} &
\colhead{RA} &
\colhead{DEC}  \\
%\colhead{Sources\tablenotemark{a}} \\
\colhead{}    & 
\colhead{}    & 
\colhead{}    & 
\colhead{}    & 
\colhead{}    & 
\colhead{}    & 
\colhead{}    & 
\colhead{}    & 
\colhead{[ppm]}    &  
\colhead{hr [J2000]}    &  
\colhead{deg [J2000]}   
%\colhead{} 
}
\startdata
\Kepa & 738 & 10358759 & 15.282 & 0.108 & 0.054 & 0.091 & 0.092 & 176 & 19 53 23.60 & +47 29 28.4 \\
\Kepb & 806 & 3832474 & 15.403 & 0.094 & 0.099 & 0.116 & 0.056 & 652 & 19 01 08.07 & +38 56 50.2  \\
\Kepc & 935 & 9347899 & 15.237 & 0.053 & 0.045 & 0.119 & 0.063 & 186 & 19 36 05.52 & +45 51 11.1 \\
\Kepd & 952 & 9787239 & 15.913 & 0.096 & 0.117 & 0.193 & 0.119 & 288  & 19 51 22.18 & +46 34 27.4
 \enddata
\tablecomments{ Information mostly from the \Kepler Input Catalog. }
\tablenotetext{a}{Contamination for each season 0-3 ( season $=$ (quarter$+2$) mod 4 ): the fractional amount of light leaking in to the target's aperture from other stars, known from the \Kepler Input Catalog.}
%\tablenotemark{b}{Combined differential photometric precision on 6 hour timescales.}
\label{tabStars}
\end{deluxetable}

\begin{deluxetable}{lcccccc} 
\tabletypesize{\scriptsize}
%\rotate
\tablecaption{Table of Stellar Properties of Hosts}
\tablewidth{0pt}
\tablehead{
\colhead{} & 
\colhead{$T_{\rm eff}$} & 
\colhead{$\log g$} & 
\colhead{$v \sin i$} & 
\colhead{[Fe/H]} & 
\colhead{$M_{\star}$} & 
\colhead{$R_{\star}$} \\
%\colhead{Sources\tablenotemark{a}} \\
\colhead{}    & 
\colhead{(K)} & 
\colhead{(cgs)} & 
\colhead{(km s$^{-1}$)}    & 
\colhead{}      & 
\colhead{$M_{\odot}$} & 
\colhead{$R_{\odot}$}
%\colhead{} 
}
\startdata
\Kepa & $5750\pm250$ (L) & $5.00\pm0.25$ (L) & $4\pm2$ (L) & $0.0\pm0.3$ (N) & 1.00$\pm$0.12 & 0.96$\pm$0.14 \\
\Kepb & $5498\pm54$ (K) & $4.77\pm0.23$ (K) & $1.94\pm0.22$ (K) &$0.18\pm0.27$ (K) & 0.99$\pm$0.08 & 0.95$\pm$0.12  \\
\Kepc & $6340\pm200$ & $4.696\pm0.300$ &  & $-0.076\pm0.400$ & 1.21$\pm$0.17 & 1.22$\pm$0.24 \\
\Kepd & $3900\pm200$ (K,M) & $4.64\pm0.30$ & & $0\pm~0.4$ & $0.58\pm0.05$ & $0.53\pm0.04$ 
 \enddata
\tablenotetext{a}{Sources for stellar properties.  Spectroscopic parameter with uncertainties indicated in parentheses are from: K=Keck Observatory, L=Lick Observatory, M=McDonald Observatory, N=NOAO.  Quoted uncertainties do not include systematic uncertainties due to stellar models. }
%\tablecomments{
% Table \ref{tabStars} is published in its entirety in the electronic edition of the {\it Astrophysical Journal Supplement}.  A portion is shown here for guidance regarding its form and content.
%} 
\label{tabStarsProps}
\end{deluxetable}
 
\clearpage
 
\begin{deluxetable}{rcccccccccl}
\tabletypesize{\scriptsize}
%\rotate
%\tablecolumns{10}
\tablecaption{Key Properties of Planets and Planet Candidates\label{tabPlanets}}
%\tablewidth{0pc}
\tablehead{
% \colhead{}  &
\colhead{$T_0$\tablenotemark{a}} & 
\colhead{$P$\tablenotemark{b}} & 
\colhead{$T_{\rm Dur}$\tablenotemark{c}} & 
\colhead{$b$ \tablenotemark{d}} & 
\colhead{$u$ \tablenotemark{d}} & 
\colhead{Depth} & 
\colhead{$R_p/R_*$\tablenotemark{e}} & 
\colhead{$R_p$\tablenotemark{f}} & 
\colhead{$a$} & 
\colhead{$M_{\rm p,max}$\tablenotemark{g}} \\ 
%\colhead{} & 
\colhead{(d)} & 
\colhead{(d)} & 
\colhead{(d)} & 
\colhead{}  & 
\colhead{}  & 
\colhead{(ppm)} & 
\colhead{}  & 
\colhead{$R_{\oplus}$} & 
\colhead{(AU)} & 
\colhead{$M_{\rm Jup}$} }
\startdata
29b \qquad 82.750$\pm$0.009 &  10.3376$\pm$0.0002 &  0.117$\pm$0.003 & 0.0 -- 0.85 & 0.0 -- 0.5 &  1204 &  0.0343$\pm$0.0008   & 3.6$\pm$0.5 & 0.09 &   0.4 \\ 
29c \qquad 78.471$\pm$0.016 &  13.2907$\pm$0.0004 & 0.127$\pm$0.006 & \nodata & \nodata &    871 &    0.0280$\pm$0.0012    & 2.9$\pm$0.4  & 0.11 &   0.3 \\ 
\hline
30b ~\qquad  \quad  83.04$\pm$0.41 &  29.329$\pm$0.022 &  0.200$\pm$0.004 & \nodata & \nodata &  1540  &  0.0351$\pm$0.0005 & 3.6$\pm$0.5  & 0.18 &    0.2 \\ 
30c \qquad 176.904$\pm$0.005 &  60.3251$\pm$0.0008 &  0.2392$\pm$0.0011 & 0.44$\pm$0.04 & 0.584$\pm$0.023  &   20578   & 0.1375$\pm$0.0011 & 14.3$\pm$1.8  & 0.30 &  9.1 \\ 
30d  \qquad  87.220$\pm$0.038 & 143.213$\pm$0.013 &  0.322$\pm$0.003 & 0.52$\pm$0.04 & 0.56$\pm$0.05  &  11014 &  0.1020$\pm$0.0014 & 10.6$\pm$1.4  & 0.5 &    17 \\ 
\hline
935.04  \qquad 85.10$\pm$0.04 & 9.6172$\pm$0.0005 &   0.176$\pm$0.014 & \nodata  &   0 -- 0.6 & 161 & 0.0136$\pm$0.0006  & 1.8$\pm$0.4  & 0.09 & \nodata  \\ 
31b \qquad  92.141$\pm$0.006 &  20.8613$\pm$0.0002 &  0.208$\pm$0.002 & 0 -- 0.75 & 0.38 -- 0.70  &  1895 &  0.0411$\pm$0.0006 & 5.5$\pm$1.1  & 0.16& \nodata \\ 
31c \qquad  74.191$\pm$0.007 &  42.6318$\pm$0.0005 & 0.251$\pm$0.004  & 0 -- 0.8 & 0.18 -- 0.67  &  1729  &   0.0400$\pm$0.0007 & 5.3$\pm$1.1  & 0.26 &   4.7 \\ 
935.03 \quad  67.942$\pm$0.009 &  87.6451$\pm$0.0014 &  0.344$\pm$0.010 & 0 -- 0.9 & 0.1 -- 1.0 &  959 &   0.0291$\pm$0.0011 & 3.9$\pm$0.8  & 0.4 &    6.8 \\ 
\hline
952.05 \qquad  65.54$\pm$0.04 & 0.74296$\pm$0.00007  &  0.039$\pm$0.004 & \nodata & \nodata &  224 & 0.0142$\pm$0.0007   & 0.82$\pm$0.07  &0.013  &  \nodata  \\ 
952.04 \qquad  66.61$\pm$0.03 &   2.8960$\pm$0.0003 &  0.053$\pm$0.004 & \nodata & \nodata &  671 & 0.0259$\pm$0.0013 & 1.5$\pm$0.1  & 0.033 & \nodata \\ 
32b \qquad  74.902$\pm$0.008 &   5.90124$\pm$0.00010 & 0.088$\pm$0.005 &\nodata  & \nodata &  1650 & 0.0389$\pm$0.0019 & 2.2$\pm$0.2  & 0.05 &   4.1 \\ 
32c \qquad  77.378$\pm$0.013 &   8.7522$\pm$0.0003 &  0.097$\pm$0.010 & \nodata &\nodata &  1453 & 0.0352$\pm$0.0033 & 2.0$\pm$0.2  & 0.09 &   0.5 \\ 
952.03 \quad  88.211$\pm$0.011 &  22.7802$\pm$0.0005 &  0.124$\pm$0.003 & \nodata & 0.0 -- 0.75 &  2181 &  0.0467$\pm$0.0011 & 2.7$\pm$0.2  & 0.13 & \nodata \\
\enddata
\tablenotetext{a}{BJD-2454900.  Rather than statistical error bars, we report the RMS of the measured transit timing deviations.}
\tablenotetext{b}{Derived from the measured transit times.  Error bar is the RMS of the measured transit timing deviations, divided by the number of transits that the data span. }
\tablenotetext{c}{Duration of time that the center of the planet overlies the stellar disk. }
\tablenotetext{d}{Values with $\pm$ error bars are from formal fits.  However, often we found $b$ and $u$ to be quite poorly constrained and/or degenerate, so we computed grids in $b$,~$u \in [0,1]$, and give either a $2$--$\sigma$ ($\Delta \chi^2=4$) range, or if this range covers the whole grid, no value at all.  }
\tablenotetext{e}{Takes into account contamination values from Table \ref{tabStars}.}
\tablenotetext{f}{Using stellar radii from Table \ref{tabStarsProps}.}
\tablenotetext{g}{Based on assumption of dynamical stability and stellar mass from Table \ref{tabStarsProps}.}

\end{deluxetable}

\clearpage
%\clearpage
\begin{deluxetable}{rrccccccccc} 
%\tabletypesize{\scriptsize} 
%\rotate 
\tablecaption{Table of Statistics for Pairs of Planets Candidates\label{tabPlanetPairs}} 
\tablewidth{0pt} 
\tablehead{ 
\colhead{KOI$_{\rm in}$} &  
\colhead{frequency$_{\rm in}$} &
\colhead{FAP$_{\rm in}$} &
\colhead{KOI$_{\rm out}$} &  
\colhead{frequency$_{\rm out}$} &
\colhead{FAP$_{\rm out}$} 
%\\ 
%\colhead{}  &  
%\colhead{}  &  
%\colhead{}  & 
%\colhead{}  & 
%\colhead{}  & 
%\colhead{}  & 
%\colhead{}  & 
%\colhead{}  &  
%\colhead{}  & 
%\colhead{}   
   } 
\startdata
 \Kepa b & 0.00026034 & \bf{0.00018} &  \Kepa c & 0.00026086 & 0.01513 \\
 \hline
 \Kepb b & 0.00098529 & $\mathbf{<10^{-5}}$ & \Kepb c & 0.00099742 & \bf{0.00016} \\
 \Kepb b & 0.00698174 & 0.70891 & \Kepb d & 0.00076809 & 0.29269 \\
 \Kepb c  & 0.00259824 & 0.00855 & \Kepb d & 0.00262702 &  $\mathbf{<10^{-5}}$ \\
 \hline
 \Kepc b  & 0.00100550 & $\mathbf{<10^{-5}}$ &    \Kepc c & 0.00100732 & 0.03360\\
 \Kepc b  & 0.01140540 & 0.53060&    935.03& 0.00229586& 0.96650\\
 \Kepc c  & 0.00064850 & 0.02580 &   935.03 &0.00064007& 0.13400\\
 \hline
 952.04 & 0.00642701 & 0.11427 &    \Kepd b & 0.00642994 & 0.27161 \\
 952.04 & 0.11678254 & 0.09436 &    \Kepd c & 0.00252686 & 0.16165 \\
 952.04 & 0.04389705 & 0.51453 &  952.03 & 0.00587133 & 0.61985 \\
 \Kepd b & 0.00390700 &  \bf{0.00001} &  \Kepd c & 0.00391415 & 0.05363 \\
 \Kepd b & 0.04389928 & 0.84870 &  952.03 & 0.00613203 & 0.49878 \\
 \Kepd c & 0.02646319 & 0.87862 &  952.03 & 0.01743588 & 0.65570 \\
 \hline
 \hline
 Kepler-9b  & 0.00066337  &  \bf{0.00002}  &   Kepler-9c &  0.00066211 &  0.00170 \\
 \hline
 Kepler-18c &  0.00373946 &  $\mathbf{<10^{-5}}$  &   Kepler-18d  & 0.00381809  & $\mathbf{<10^{-5}}$ \\
 \hline
 168.02 & 0.02919133 & 0.80572 &  Kepler-23b  & 0.02919325 & 0.80219 \\
 168.02 & 0.01021819 & 0.37630 &  Kepler-23c  & 0.01022921 & 0.65446 \\
 Kepler-23b  & 0.00215313 &  \bf{0.00084} &  Kepler-23c  & 0.00214191 &  $\mathbf{<10^{-5}}$ \\
 \hline
1102.04 & 0.00994495 & 0.22250  & 1102.02  & 0.00994282 & 0.35748\\
1102.04 & 0.07345063 & 0.75543  & 1102.01 & 0.00762697 & 0.91916\\
1102.04 & 0.05264294 & 0.25841  & 1102.03 & 0.02506453 & 0.56294\\
Kepler-24b & 0.00237619 & $\mathbf{<10^{-5}}$  & Kepler-24c & 0.00237709 & $\mathbf{<10^{-5}}$\\
Kepler-24b & 0.01749593 & 0.85933  & 1102.03 & 0.01752736 & 0.14017\\
Kepler-24c & 0.00424632 & 0.63164  & 1102.03 & 0.00424409 & 0.57523 \\
 \hline
 Kepler-25b & 0.00305506 &  $\mathbf{<10^{-5}}$ &  Kepler-25c & 0.00306491 & 0.02136 \\
 \hline
 250.03 & 0.11935114 & 0.02223 &   Kepler-26b & 0.03794286 & 0.84931 \\
 250.03 & 0.05797149 & 0.06283 &   Kepler-26c & 0.00765104 & 0.10796 \\
 Kepler-26b & 0.01119109 & 0.05656 &   Kepler-26c & 0.01118508 & 0.28181 \\
 \hline
 Kepler-27b & 0.00137000 &  \bf{0.00061} &  Kepler-27c & 0.00137385 &  \bf{0.00083} \\
 \hline
 Kepler-28b & 0.00426232 &  $\mathbf{<10^{-5}}$ & Kepler-28c & 0.00429017 & 0.00239 
\enddata
\tablecomments{
Theoretically predicted O-C frequencies are in cycles per day.  False Alarm Probabilities (FAP) are bold, if the detection is considered significant ($FAP<10^{-3}$).  Systems above the line are presented as planetary system discoveries in this paper.   Systems below the double horizontal line are presented as planetary system discoveries in \cite{2010Holman}, in \cite{2011Cochran}, and in the companion papers, papers II and III. 
% Table \ref{tabPlanets} is published in its entirety in the electronic edition of the {\it Astrophysical Journal Supplement}.  A portion is shown here for guidance regarding its form and content.
}
\end{deluxetable}

\clearpage
\begin{deluxetable}{ccccccccc} 
\tabletypesize{\scriptsize} 
%\rotate 
\tablecaption{Example TTV Solution for Planets Candidates} 
\tablewidth{0pt} 
\tablehead{ 
\colhead{system} &  
\colhead{planet} &  
\colhead{ $P$ } &
\colhead{$T_0$ } &
\colhead{$e \cos \omega$} &  
\colhead{$e \sin \omega$} &
\colhead{$M_p$} &
\colhead{$M_p$ (3-$\sigma$ upper- } &
\colhead{$\chi^2$ / d.o.f.} 
\\ 
\colhead{$M_\star$ ($M_\odot$)}  &  
\colhead{}  &  
\colhead{(days)}  &  
\colhead{BJD-2454900}  & 
\colhead{}  & 
\colhead{}  & 
\colhead{(formal, $M_\oplus$)} &
\colhead{limit, $M_\oplus$)} &
\colhead{}  
} 
\startdata
 Kepler-23& b &     7.107977 & 320.024941 &   0.030090 &  -0.061653 &     4.8  & 80 &112.1 / 60\\
  1.21 &    &   $\pm$ 0.000934 &  $\pm$0.007984 &  $\pm$0.454676  & $\pm$0.357376 & $\pm$ 15.6& & \\
  &  c    &    10.741598 & 324.116640 &   0.034012 &  -0.056846 &    15.0  &  700 & 38.2 / 39  \\
   &    &   $\pm$ 0.001250 &  $\pm$0.002704 & $\pm$0.386216 &  $\pm$0.294888 &   $\pm$49.8 & & \\
\hline
 Kepler-24 & b   &     8.163886 & 326.108132 &  -0.336422 &   0.238848 &    56.1 & 130 &  86.3 / 65 \\
1.10    &   &   $\pm$ 0.004031 &  $\pm$0.005328 &  $\pm$0.090501  & $\pm$0.117101 & $\pm$ 15.8& & \\
    & c   &    12.315302 & 329.547776 &  -0.267037 &   0.202843 &   102.8  & 350 &  131.7 / 40\\
    &   &   $\pm$ 0.003671 &  $\pm$0.006364 & $\pm$0.082641 &  $\pm$0.099609 &   $\pm$21.4 & & \\
\hline
 Kepler-25 & b &     6.238383 & 323.057366 &  -0.011764 &  -0.007540 &     8.1  &  310 &83.4 / 69  \\
 1.10  &    &   $\pm$ 0.000100 &  $\pm$0.000547 &  $\pm$0.002010  & $\pm$0.003149 & $\pm$  3.1 &  &  \\
   & c &    12.720640 & 327.772432 &  -0.029885 &  -0.020024 &    13.3  & 30 &43.9 / 31\\
    &   &   $\pm$ 0.000179 &  $\pm$0.000337 & $\pm$0.006415 &  $\pm$0.003600 &   $\pm$ 3.9  &  &  \\
\hline
Kepler-26 & b &    12.281204 & 324.486687 &  -0.009929 &  -0.360464 &     2.0  & 4.6 &65.1 / 30\\
0.55    &   &   $\pm$ 0.000580 &  $\pm$0.000982 &  $\pm$0.018073  & $\pm$0.366736 & $\pm$  1.0 &  &  \\
 & c &    17.253059 & 324.406889 &  -0.017962 &  -0.317353 &     3.9  & 20 &25.4 / 19\\
   &    &   $\pm$ 0.001135 &  $\pm$0.001088 & $\pm$0.015168 &  $\pm$0.316959 &   $\pm$ 1.4  &  &  \\
\hline
 Kepler-27 & b &    15.337102 & 321.703954 &   0.015924 &   0.000148 &    28.5  & 800 & 34.3 / 25\\
 0.94  &    &   $\pm$ 0.002742 &  $\pm$0.001715 &  $\pm$0.010802  & $\pm$0.001418 & $\pm$ 12.6 &  &  \\
  & c &    31.330788 & 337.066843 &   0.032298 &   0.003735 &    51.4  & 260 & 20.2 / 10 \\
   &    &   $\pm$ 0.000717 &  $\pm$0.001887 & $\pm$0.026605 &  $\pm$0.009598 &   $\pm$43.7  &  &  \\
\hline
 Kepler-28 & b &     5.911764 & 323.929704 &  -0.050253 &  -0.075099 &     3.8  & 320 &  54.4 / 70 \\
0.89  &     &   $\pm$ 0.000426 &  $\pm$0.002061 &  $\pm$0.175120  & $\pm$0.469131 & $\pm$  6.9 &  &  \\
   &  c &     8.986416 & 324.352707 &  -0.020126 &  -0.078354 &     4.9  & 50 & 93.3 / 45 \\
    &   &   $\pm$ 0.000612 &  $\pm$0.002432 & $\pm$0.139930 &  $\pm$0.376697 &   $\pm$ 9.3  &  &  \\
\hline
   Kepler-29  & b    &     10.335908 &  320.506863 &   -0.049991&    -0.039249  &     2.8  & 8.2  & 42.5 / 41\\
1.05 &  &   $\pm$ 0.001440   &$\pm$  0.002675   & $\pm$ 0.035113   & $\pm$  0.033556  &   $\pm$2.4  &    &\\
&  c     &    13.293451 &  331.005742   & -0.008036  & -0.050000   &      2.3 &  5.4  & 54.4 / 29\\
 &  &    $\pm$ 0.001333  &  $\pm$  0.003308    & $\pm$ 0.014426   & $\pm$ 0.000000    &   $\pm$2.0&    & \\
\hline
  Kepler-30  &  b &      29.221117 &    346.362257 &      0.112642  &     0.058682  &  3.3 &  7.3 & 19.5 / 10 \\
 0.99        &      &  $\pm$0.009642    &  $\pm$0.012845    &  $\pm$0.025241    &  $\pm$0.031588       &  $\pm$1.7 & & \\
            &  c & 60.326939  & 357.882515    & 0.036164   & -0.008941    &  240 & 870 & 10.5 / 2 \\
        &     &  $\pm$0.001034    &  $\pm$0.000966   &  $\pm$ 0.019612   &  $\pm$ 0.003552      &  $\pm$ 90 & & \\
          &  d  &      143.337469  & 373.644127    &  -0.018594   & -0.000196     &  22 & 160 & 0.033 / 0 \\
        &     &  $\pm$   0.040298    &  $\pm$ 0.013796   &  $\pm$  0.019412    &  $\pm$ 0.010622   &  $\pm$ 14 & & \\
\hline
 Kepler-31& b   &   20.856113  & 321.610535  &  0.007721 &  -0.000270  &  18.4   & 140 &  27.8 / 25 \\
 1.09 &    &   $\pm$  0.002883&     $\pm$ 0.002182   &  $\pm$ 0.003806 &    $\pm$ 0.005266   &    35.5   &  &  \\ 
& c   & 42.637618  &  329.984885  & -0.001835   & 0.031462  &  34.9   & 120 & 13.9 / 8  \\
  &  &   $\pm$ 0.009447  &   $\pm$ 0.002964  &   $\pm$ 0.015429 &    $\pm$ 0.034318  &  21.2  &  & \\
\hline
 Kepler-32 & b   &     5.900841 & 322.747894 &  -0.003717 &  -0.000888 &     7.2  & 24 &97.7 / 71  \\
 0.49   &   &   $\pm$ 0.000188 &  $\pm$0.001616 &  $\pm$0.008006  & $\pm$0.001491 & $\pm$  4.1 &  &  \\
 & c  &     8.752819 & 322.449605 &   0.001713 &   0.001658 &     5.2  & 18 &70.8 / 47  \\
   &    &   $\pm$ 0.000545 &  $\pm$0.002362 & $\pm$0.002330 &  $\pm$0.001150 &   $\pm$ 3.5  &  & 
\enddata
\tablecomments{
A sample of TTV-fit results for each system presented here. 
}
\label{tabSolutions}
\end{deluxetable}

\clearpage

\begin{deluxetable}{lcccc}
\tablecaption{Transit Times for Kepler Transiting Planet Candidates\label{tabTTs}}
\tablehead{
\colhead{KOI} & \colhead{ n } & \colhead{ $t_n$}    &  \colhead{TTV$_n$} & \colhead{ $\sigma_n$} \\ 
   \colhead{}  &   \colhead{}     &  \colhead{BJD-2454900} &  \colhead{(d)}  &  \colhead{(d)} }
\startdata
 738.01 & \multicolumn{4}{c}{$        82.749505 + n \times        10.337583$} \\
 738.01 &            0 &   82.7642 &  0.0147 & 0.0081 \\ 
 738.01 &            1 &   93.1033 &  0.0162 & 0.0074   %\end{longtable}
 \enddata
\tablecomments{Table \ref{tabTTs} is published in its entirety in the 
electronic edition of the {\it Astrophysical Journal} or from the 
``tttable'' within the source files of this arxiv posting.  A portion is 
shown here for guidance regarding its form and content.}
\end{deluxetable}

\bibliography{ttvconfirm} \bibliographystyle{apj}

%\clearpage
%\input{figures}

%\clearpage

\end{document}